\documentclass{aa}
   \usepackage{graphicx}
   \begin{document}
   \title{Flux evolution of superluminal components in blazar 3C454.3} 
   \author{S.J.~Qian\inst{1}} 
   \institute{National Astronomical Observatories, Chinese Academy of Sciences,
    Beijing 100012, China, \it{rqsj@bao.ac.cn}}
  \date{Compiled by using A\&A-latex}
  \abstract{The kinematic behavior of superluminal components observed at 
   43GHz in blazar 3C454.3 were model-fitted with their
    light curves interpreted in terms of their Doppler-boosting
    effect.}  {The relation
    between the flux evolution of superluminal components and their 
    accelerated/decelerated motion or the increase/decrease in their
    Lorentz/Doppler factor was investigated.}
      {The  precessing jet-nozzle scenario previously proposed by Qian et al. 
   (1991, 2018a, 2021) and Qian (2018b, 2022a, 2022b) was applied to 
   consistently model-fit the kinematic behavior and light curves for
     two superluminal
    components (B4 and B6) measured by Jorstad et al. (2005).}{For both B4
     and B6 which were ascribed respectively to the jet-A
    and jet-B of the double-jet structure assumed for 3C454.3, their
     kinematic features were well model-fitted 
    with their bulk Lorentz factor and Doppler factor (as function of time)
     convincingly derived. It is shown that the light curves of the radio 
    bursts associated with knot B4 and knot B6 can be well explained in terms
    of their Doppler-boosting effect. Similarly, for the knot R3 observed 
    at 15\,GHz (Qian et al. 2014, Britzen et al. 2013) the interpretation 
    of its kinematic behavior and  light curve is presented in the appendix.}
    {We emphasize that the interpretation of the flux evolution of superluminal
    components combined with the mode-fit of their kinematics is important
    and fruitful. This kind of combined investigation not only can greatly 
   improve the model-simulation of their kinematics with properly selecting 
   the model-parameters (especially the bulk-Lorentz factor and Doppler factor 
   as functions of time), but also their light curves can be well interpreted
    in terms of their Doppler-boosting effect. Therefore, we can almost 
   completely (or perfectly) understand the physical nature of these
  components: their kinematic/dynamic characteristics and emission properties.}
    \maketitle
  \section{Introduction}
    3C454.3 (redshift z=0.859) is one of the most prominent blazars, radiating 
   across the entire electromagnetic spectrum 
  from radio/mm through IR/optical/UV and X-ray to high-energy
  $\gamma$-rays with strong variability at all the wavebands on various 
   timescales. For example, in May 2005 its optical flaring activity reached
   an unprecedented level with the R-band magnitude $m_R$$\sim$12\,mag,
   an extermely bright state (Raiteri et al. \cite{Ra07}, Villata et al.
   \cite{Vi06}). During the time-interval 2007--2010 3C454.3
    underwent an  exceptionally strong $\gamma$-ray activity period
   (Vercellone et al. \cite{Ve09}, \cite{Ve11}):
   it was the brightest $\gamma$-ray  source in the sky on 2009 December 2-3
    and 2010 November 20, a factor of $\sim$2 and $\sim$6 brighter than the
    Vela pulsar (Vercellone et al. \cite{Ve10}),
    respectively. Particularly important, both $\gamma$-ray flares
    were associated with the ejection of superluminal components  observed 
    by VLBI-observations (Jorstad et al. \cite{Jo13}).\\
      3C454.3 is an extremely variable flat-spectrum radio quasar with 
   superluminal components steadily ejected from its radio core.
   Based on the 1981--1986
    VLBI observations at centimeter wavelengths Pauliny-Toth et al. 
   (\cite{Pau87}) firstly detected some distinctive features of its 
   superluminal components: superluminal brightening of stationary structure,
   apparent acceleration of superluminal components and extreme curvature in
   the apparent trajectory.\\
    Multiwavelength monitoring campaigns on 3C454.3 have been performed during
  $\gamma$-ray outbursts to investigate the correlation between radio, optical,
   X-ray and $\gamma$-ray flares, especially the correlation between the
   $\gamma$-ray outbursts and the ejection of superluminal components from the
     radio core on parsec-scales (e.g., Jorstad et al. \cite{Jo01},
    \cite{Jo10}, \cite{Jo13}; Vercellone et al. \cite{Ve10}). 
    These studies have greatly improved the understanding of the
    disinctive high-energy phenomena occurred in 3C454.3 and in other 
   $\gamma$-ray blazars.\\ 
    Moreover, VLBI observations have revealed more peculiar features in its
   morphological structure and kinematics.  For example,\\
   \\
    (1) at 15\,GHz 
   an arc-like structure around the core was detected (Britzen et al. 
   \cite{Br13}; cf. Fig.A.1) in an area delimited by core distance
    [1.5\,mas, 3.5\,mas] and position angle [--$40^{\circ}$, --$110^{\circ}$],
    which was formed by the distributed superluminal
   components. It expanded with superluminal velocities, 
   dominating the  pc-structure over $\sim$14 years.  A similar arc-like 
   structure was also observed at 43\,GHz (cf. Fig.1 below; Jorstad et al. 
   \cite{Jo05});\\
   (2) VLBI observations at 43\,GHz showed that the trajectories
   of moving components could be separated into two groups: e.g., beyond core
    distance $\sim$0.5\,mas, knot B6 (and B3) moved northwest, while knot B4 
    moved southwest (Jorstad et al. \cite{Jo05});\\
    (3) However, in a striking contrast, near the core (within core distance 
    of $\sim$0.2\,mas) knot B6 moved along an extremely curved path with its
    ejection position angle $\sim$\,--$150^{\circ}$ (at ejection epoch 1999.80),
    while knot B4 moved along a track with its ejection position angle 
    $\sim$\,--$80^{\circ}$ (at ejection epoch 1998.36). This position angle
    swing (${\sim}50^{\circ}$/yr; Jorstad et al. \cite{Jo05}) seems too fast
    to be explained in terms of a "sudden jump" in the jet-nozzle direction 
    and it could be presumed as a clue for a double-jet structure 
    in 3C454.3 with B4 and B6 being ejected from the respective jets
     (similar rapid position angle swings were observed in blazars OJ287
     and 3C279, cf. Qian \cite{Qi18b}, Qian et al. \cite{Qi19}); \\
    (4) A detailed analysis of the VLBI-kinematics observed at 43\,GHz 
    revealed some recurrent trajectory patterns: e.g., the knot pair B4/K09
    (in jet-A) and  the knot-pairs B6/K10 and B2/K16 (in jet-B) with time
    intervals of $\sim$1--2 precession periods (Qian \cite{Qi21}). This
    discovery seems significant for understanding the nature of their
    kinematics. That is, the recurrence of these regular trajectory patterns
    may not only imply some periodic behavior induced by the jet-nozzle 
    precession, but also the possible existence of 
     some common precessing trajectory pattern(s) as suggested in Qian et al.\\
    (5) The periodicity analysis of the secular optical (B-band) light curve
    for 3C454.3 by using Jurkevich method (Su \cite{Su01}) resulted in a period 
    of $\sim$12.4\,yr. Similarly, an analysis of its light curves at
     multifrequencies
     (4.8, 8, 14.5, 22 and 37\,GHz) (Kudryavtseva \& Pyatunina, \cite{Ku06})
     revealed the periodicity in its flux variations with a period of
     12.4 years also. Based on the quasi-regular double-bump structure
    in its 4.8 and 8\,GHz  light curves, Qian et al. (\cite{Qi07})
     proposed a binary black hole model 
     with a double-jet structure to explain these light curves;\\
     (6) VLBI observations at 15\,GHz detected a radio burst 
     (during 2005--2011) for superluminal component R3 associated with  an 
     extreme curvature in its trajectory in the 
      outer jet regions (at core distances $r_n$$\sim$2--3.5\,mas).\\
    \\
     In this paper we shall investigate  the flux evolution observed at 
     43\,GHz for two superluminal components in 3C454.3 (knot B4 and knot B6)
     and  the connection with their kinematic properties. In order to search
     for the association of the flux variations with their Doppler-boosting
     effect, the model-fitting of their kinematic behaviors were performed
     more closely, taking into consideration of more details in the curves
     delineating their kinematics (e.g., the observed details in their core
     distance $r_n(t)$ and coordinate $Z_n(t)$ as function of time), which were 
     ignored in the previous studies.\\
     3C454.3 has a very complex structure at 43\,GHz. A map observed at
     2001.28 is shown in Figure 1 (cf. the sequence of maps presented in 
     Fig.15 in Jorstad et al. \cite{Jo05}).\\
     \\
     We shall utilize the results obtained for 3C454.3 by Qian et al.
     (\cite{Qi21}). In that work its thirteen superluminal components 
     observed at 43\,GHz were
     separated into two groups: group-A including the six components (B4,
     B5, K2, K3, K09 and K14) and group-B including the seven components (B1,
     B2, B3, B6, K1, K10 and K16).\footnote{As identified in Jorstad 
     et al. (\cite{Jo05}, \cite{Jo13}).}. Moreover, a double-jet structure
     (jet-A plus jet-B) was assumed to eject the superluminal knots
     of group-A and group-B, respectively. Interestingly, the kinematical 
     behavior of the superluminal knots ascribed to group-A and group-B could
     be well model-fitted respectively in the framework of our precessing 
     jet-nozzle scenario. It was found that both jets precess
     with the same precession period of 10.5\,yr, but have different precessing
     common trajectory patterns. \\
      As a supplement we shall discuss the flux evolution associated with the 
      Doppler-boosting effect for the superluminal knot R3 observed at 15\,GHz.
      \begin{figure*}
      \centering
      \includegraphics[width=10cm,angle=0]{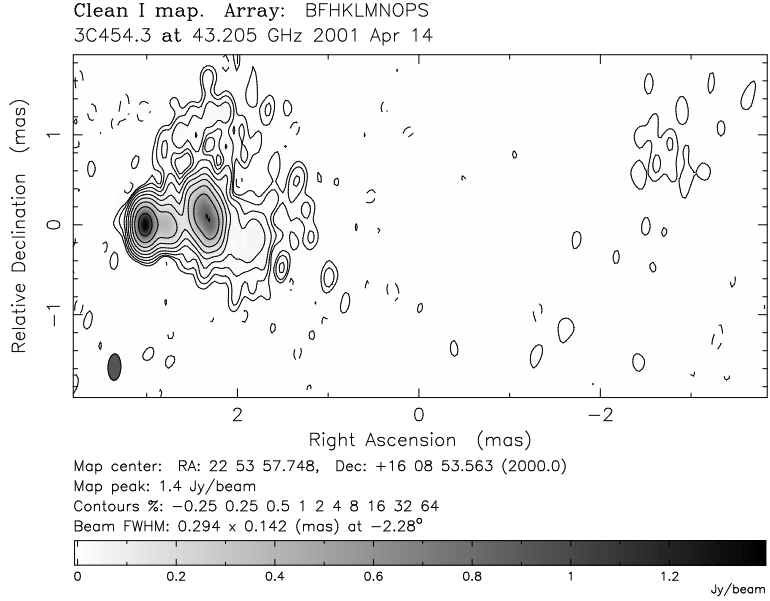}
      \caption{A 43\,GHz image of 3C454.3 observed at 2001.28 (by courtesy 
      of S.G.Jorstad, cf. Fig.44 in Jorstad et al. \cite{Jo05}; also cf. 
      Fig.15 (for a sequence of total intensity images during 1998.17--2001.28)
      and Fig.18 (for the trajectories of B4 and B6). On this map the positions
      of three knots (B4, B6 and D) should be
      indicated: knot B4 at position $[r_n,PA]$=[1.16\,mas, --$92.5^{\circ}$],
      knot B6 at position $[r_n,PA]$=[0.70\,mas, --$80.1^{\circ}$] and knot D
      at position $[r_n,PA]$=[6\,mas, --$82^{\circ}$]. The ejection position 
     angles of B4 and B6 were $-80^{\circ}$ and $-150^{\circ}$ respectively.
     Their ejection epochs were 1998.36 and 1999.80. Thus a  position angle
     swing of $\sim70^{\circ}$ occurred in a short time-interval of
     $\sim$1.3 years. It seems that such a rapid swing could not be induced 
     by a sudden "jump" in the position angle of the jet-nozzle, but could be
      resulted from the two knots being ejected by two different nozzles.
      In the precessing jet-nozzle scenario for 3C454.3 (Qian et al. 
     \cite{Qi21}), knot B4 and knot B6 are ascribed to jet-A and jet-B, 
     respectively. In addition, there is an arc-like structure around the core
     which distributed in the area delimited by $r_n{\simeq}$[1.0--1.5\,mas]
     and PA$\simeq$[--$30^{\circ}$, --$110^{\circ}$]. This arc-like structure 
     is quite similar to that  detected at 15\,GHz by Britzen
      et al. (\cite{Br13}); cf. Fig.A.1 in the appendix.}
     \end{figure*} 
    \begin{figure*}
  \includegraphics[width=9cm,angle=90]{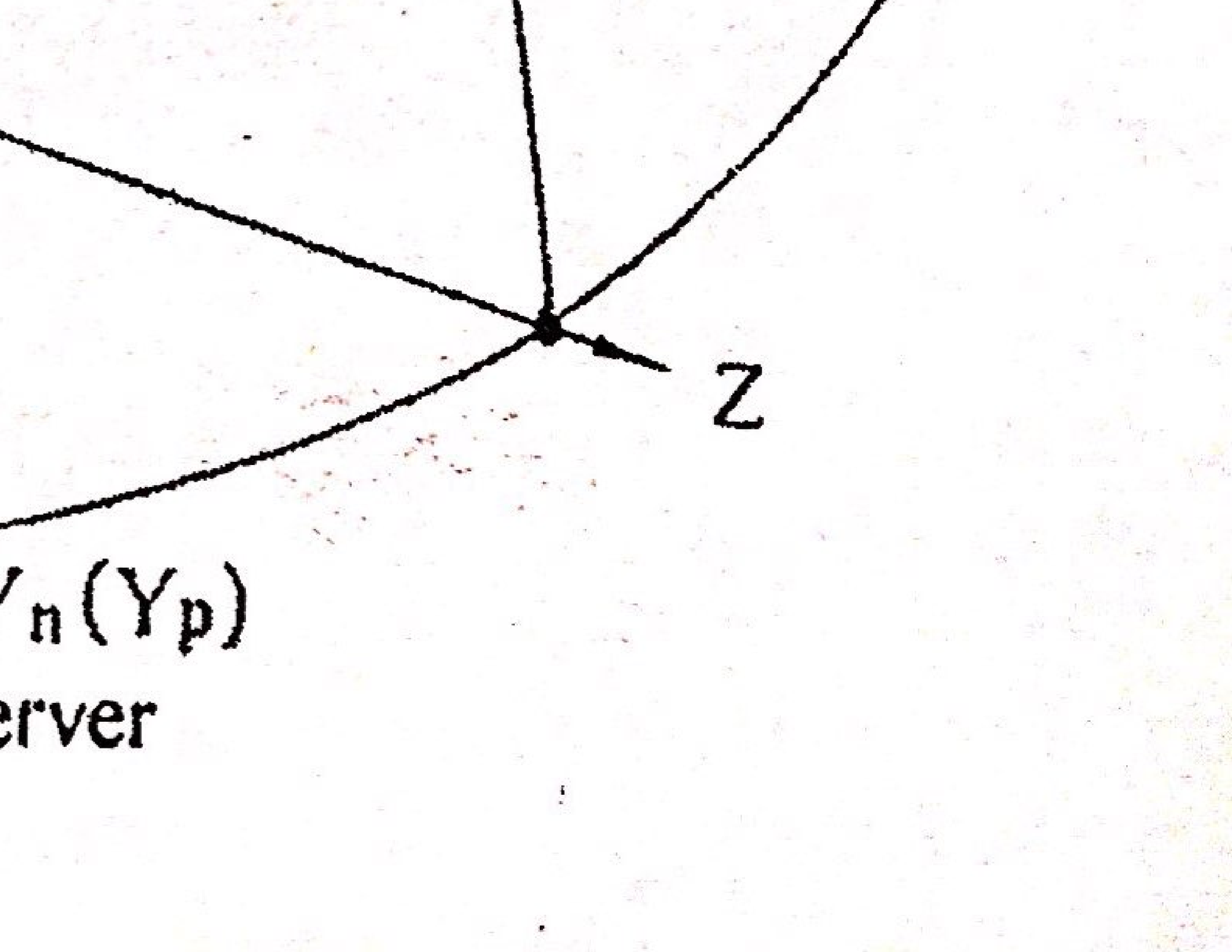} 
  \centering
  \caption{Geometry of the precessing nozzle scenario with a helical 
   trajectory pattern.}
  \end{figure*}
  \section{Recapitulation of the precessing nozzle model:
   Geometry and formalism} 
    In order to investigate the kinematic behavior and distribution of 
   trajectory of the superluminal components on parsec scales in terms of our
   precessing jet-nozzle scenario, we define a 
   special geometry in general case, where four coordinate systems are
   introduced (Qian et al. \cite{Qi21}; Qian \cite{Qi22a}, \cite{Qi22b}), as 
   shown in Figure 2: (X,Y,Z), ($X_n,Y_n,Z_n$), ($X_p,Y_p,Z_p$), 
    and  (${\it{x}}'$,${\it{y}}'$,${\it{z}}'$).
     Z-axis directs toward the observer,
    ($X_n,Y_n$) and ($X_p,Y_p$) define the plane of  the sky with $X_n$-axis
    pointing toward the negative right ascension and $Z_n$-axis toward
    the north pole. Parameter $\epsilon$  defines the angle between Z-axis
   and $Y_n$-axis and  $\psi$ the angle between X-axis and $X_n$-axis. Thus
   parameters $\epsilon$ and $\psi$ define the plane where the jet-axis 
   ($x_0(z_0)$) locates (see below).  The helical trajectory
   of a knot can be described by the parameters A($s_0$) and $\phi(s_0)$, which 
   are defined in the coordinate system (${\it{x}}'$,${\it{y}}'$,${\it{z}}'$)
   , where $s_0$ is the arc-length along the axis of the helix and 
   ${\it{z}}'$-axis is along the tangent of the axis of the helix. In addition, 
    The helical trajectory precesses around the jet axis, producing the 
   trajectories of different knots ejected at different times. That is 
   superluminal components are ejected from the precessing jet nozzle at 
    the helical phase $\phi$ which is related to the precession phase
   $\phi_0$. $\phi_0$ varies over a range of [0, 2$\pi$] during a 
   precession period and is related to its ejection epoch $t_0$. \\
    In the following we will
   adopt the formalism  of the precessing nozzle scenario given in 
   Qian et al. (\cite{Qi21}).
   The projection of the spatial trajectory of a knot on the sky-plane can 
    be calculated by using the transformation between the coordinate systems.
   Superluminal knots are assumed to move on a helical trajectory defined by 
   the parameters A($s_0$) (amplitude) and  $\phi(s_0)$ (phase), 
   $s_0$ -- arc-length along the jet axis which is defined by
   \begin{equation}
       {x_0}(z_0)=p(z_0)({z_0}^{\zeta})
   \end{equation}
    where
   \begin{equation}
    p({z_0})={p_1}+{p_2}[1+exp(\frac{{z_t}-{z_0}}{{z_m}})]^{-1}
   \end{equation}
   $\zeta$, $p_1$, $p_2$, $z_t$ and $z_m$ are constants. $\zeta$=2 represents
    a parabolic shape for the axis of the helical trajectory.\\ 
   The arc-length along the helical trajectory is $s_0$:
   \begin{equation}
   {s_0}=\int_{0}^{z_0}{\sqrt{1+(\frac{d{x_0}}{d{z_0}})^2}}{d{z_0}}
   \end{equation}
   The helical trajectory of a knot can be described in (X,Y,Z) system as 
   follows.
   \begin{equation}
    X({s_0})=A({s_0}){\cos{\phi({s_0})}}{\cos{\eta({s_0})}}+{x_0}
   \end{equation}
   \begin{equation}
    Y({s_0})=A({s_0}){\sin{\phi({s_0})}}
   \end{equation}
   \begin{equation}
    Z({s_0})=A({s_0}){\cos{\phi({s_0})}}{\sin{\eta({s_0})}}+{z_0}
   \end{equation}
   where $\tan{\eta({s_0})}$=$\frac{d{x_0}}{d{z_0}}$. The projection of the
    helical trajectory on the plane of the sky (or the apparent trajectory)
    is represented by
    \begin{equation}
        {X_n}={X_p}{\cos{\psi}}-{Z_p}{\sin{\psi}}
    \end{equation}
    \begin{equation}
        {Y_n}={X_p}{\sin{\psi}}+{Z_p}{\cos{\psi}}
    \end{equation}
    where
    \begin{equation}
        {X_p}={X({s_0})}
    \end{equation}
    \begin{equation}
        {Z_p}={Z({s_0})}{\sin{\epsilon}}-{Y({s_0})}{\cos{\epsilon}}
    \end{equation}
    All coordinates and amplitude (A) are measured in units of milliarcsecond.
    Introducing the following functions
    \begin{equation}
     {\Delta}=\arctan[(\frac{dX}{dZ})^2+(\frac{dY}{dZ})^2]^{-\frac{1}{2}}
    \end{equation}
    \begin{equation}
     {{\Delta}_p}=\arctan(\frac{dY}{dZ})
    \end{equation}
    \begin{equation}
     {{\Delta}_s}=\arccos[(\frac{dX}{d{s_0}})^2+(\frac{dY}{d{s_0}})^2+
                  (\frac{dZ}{d{s_0}})^2]^{-\frac{1}{2}}
    \end{equation}
    We can then calculate the viewing angle $\theta$, apparent transverse 
    velocity ${\beta}_{app}$, Doppler factor $\delta$ and the elapsed time T,
    at which the knot reaches distance $z_0$, as follows:
     \begin{equation}
     {\theta}=\arccos[{\cos{\epsilon}}(\cos{\Delta}+
               \sin{\epsilon}\tan{{\Delta}_p})]
     \end{equation}
     \begin{equation}
     {\delta}=[{\Gamma}(1-{\beta}{\cos{\theta}})]^{-1}
     \end{equation}
     where $\Gamma$=$[1-{\beta}^2]^{-\frac{1}{2}}$ -- bulk Lorentz factor,
     $\beta$=$\it{v}$/c,  $\it{v}$ -- velocity of the knot.
     \begin{equation}
      {{\beta}_{app}}={{\beta}{\sin{\theta}}/(1-{\beta}{\cos{\theta}})}
     \end{equation}
     \begin{equation}
     T=\int^{{s_0}}_{0}{\frac{(1+z)}{{\Gamma\delta}{v}{\cos{{\Delta}_s}}}}
                                   {d{s_0}}
     \end{equation}
    In this paper we shall adopt the concordant cosmological model 
   (${\Lambda}CDM$) with $\Omega_m$=0.3, $\Omega_{\Lambda}$=0.7, and Hubble
   constant $H_0$=70km${s^{-1}}{Mpc^{-1}}$ (Spergel et al. \cite{Sp03}). Thus 
   for 3C454.3, z=0.859, its luminosity distance  is $D_l$=5.483\,Gpc 
   (Hogg \cite{Ho99}, Pen \cite{Pe99}) and angular diameter distance $D_a$=
   1.586\,Gpc. Angular scale 1\,mas=7.69\,pc and proper motion of 1\,mas/yr
   is equivalent to an apparent velocity of 46.48\,c.
     \section{Knot B4: Model-fitting of kinematic behavior and 43\,GHz light
     curve}
     As shown in the previous work (Qian et al. \cite{Qi21}) 
    knot B4 was ejected from the nozzle associated with the jet-A 
    in our precessing jet-nozzle 
    scenario.  Other knots attributed to jet-A are B5, K2, K3, K09 and K10.
    (Jorstad et al. \cite{Jo05}, \cite{Jo10}, \cite{Jo13}). The kinematic 
    behavior of these knots has been consistently interpreted 
    in terms of our scenario (Qian et al. \cite{Qi21}), but untill now their
    flux evolution has not been investigated.\\
    Knot B4 produced two radio bursts. In order 
    to investigate their flux evolution associated with its kinematic behavior
    we adopt the model parameters for jet-A  as 
    before:\\
    \begin{figure*}
    \centering
    \includegraphics[width=5.6cm,angle=-90]{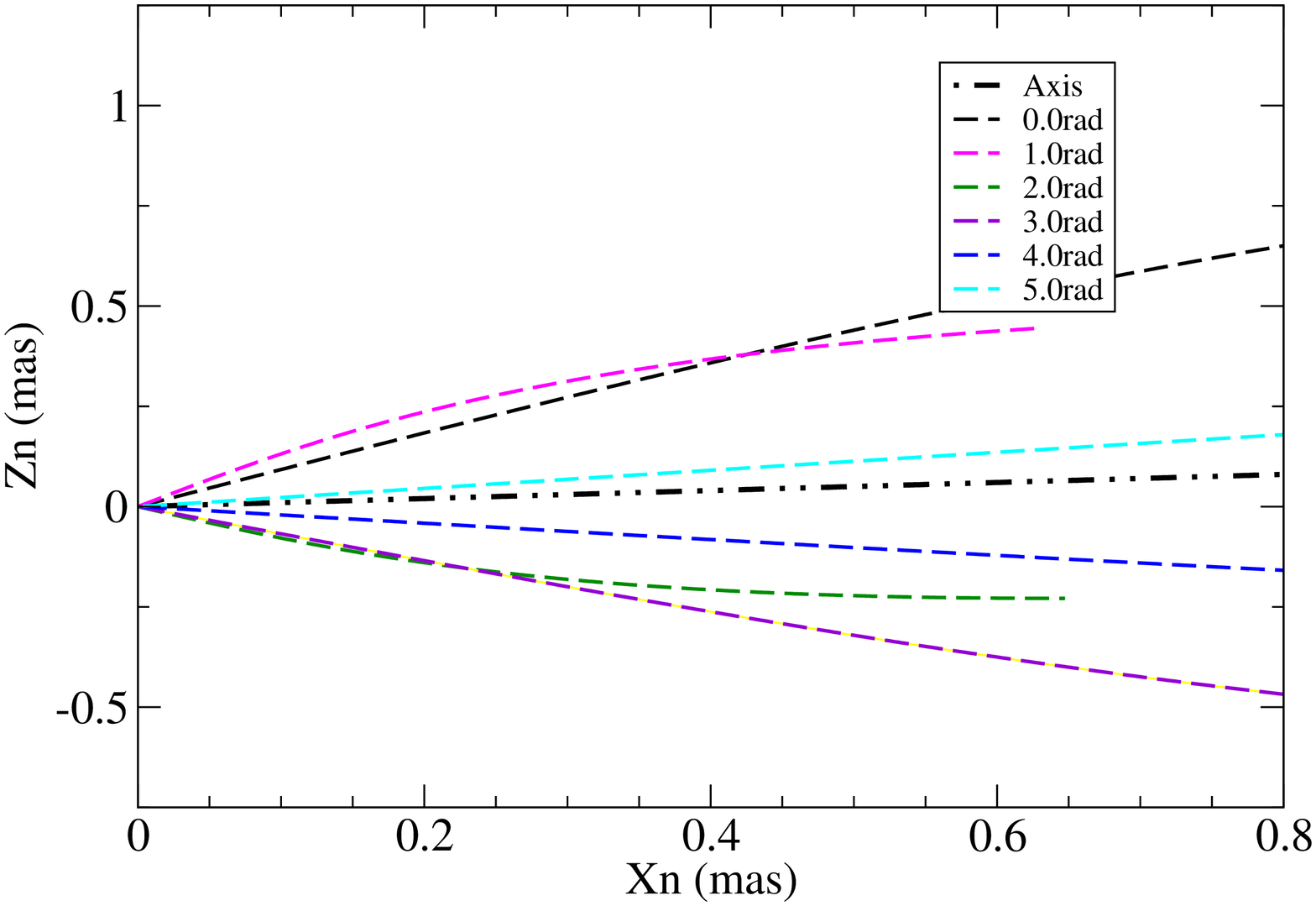}
    \includegraphics[width=5.6cm,angle=-90]{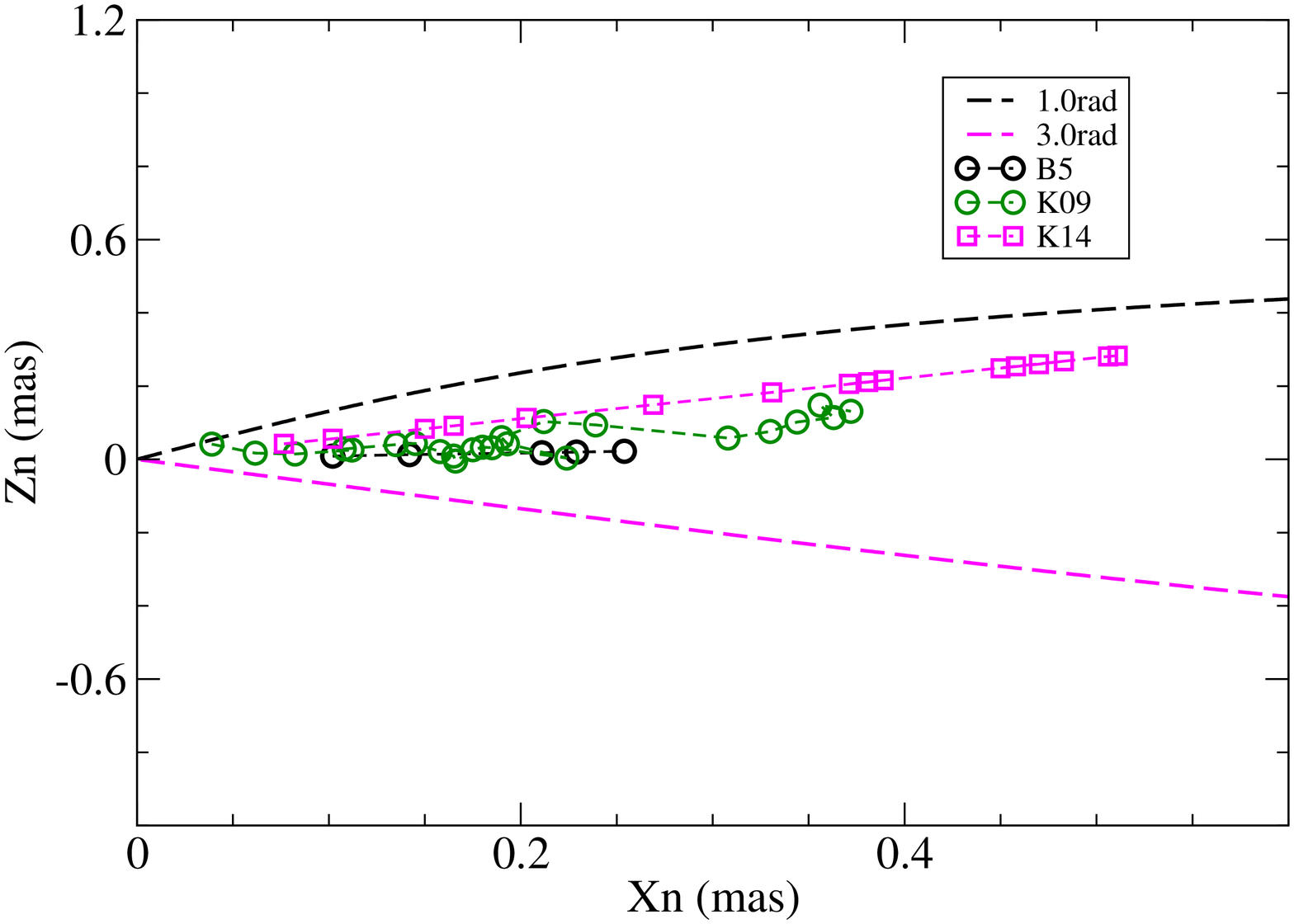}
    \caption{Jet-A. Left panel: the modeled distribution of the precessing
    common trajectories for the superluminal components at precessing phases 
    $\phi$=0.0, 1.0,2.0, 3.0, 4.0 and 5.0, respectively. Right panel: the 
    trajectories observed for knots B5, K09 and K14 within the jet-cone.
     The jet-axis is at position angle of $-84.3^{\circ}$.}
    \end{figure*}
     (a) The plane, in which the jet axis locates, is defined by 
     parameters $\epsilon$ and $\psi$: 
   $\epsilon$=0.0126\,rad=$0.72^{\circ}$, $\psi$=--0.1\,rad=--$5.73^{\circ}$.\\
     (b) The jet axis is defined by a set of parameters: $\zeta$=2, 
    $p_1$=$p_2$=0, $z_t$=33\,mas, $z_m$=3.0\,mas (cf. equations 1 and 2).\\ 
   (c) The precessing common trajectory pattern is assumed as: amplitude 
   A=$A_0$$\sin(\pi{z_0}/{Z_1})$ with ($A_0$=0.727, $Z_1$=240\,mas)
     and  the helical phase $\phi$ identically equal to the precession phase
    $\phi_0$.\\
     For jet-A the precession phase $\phi_0$ of the superluminal knots is
     related  to the ejection epoch $t_0$ as follows: 
    \begin{equation}
      {\phi_0}(rad)=4.58+\frac{2\pi}{T_0}(t_0-1998.24)
    \end{equation}   
    In order to investigate the flux evolution associated with the Doppler
     boosting effect the observed flux density will be calculated as follows:
     \begin{equation}
      {S_{obs}(\nu,t)}={S_{int}(\nu,t)}{\times}{[\delta(t)]^{3+\alpha(\nu,t)}}
     \end{equation}
     where $S_{obs}$ -- observed flux density, $S_{int}$ -- intrinsic flux 
    density ($S_{\nu}$\,$\propto$\,${\nu}^{-\alpha}$) and $\alpha$ -- spectral 
    index.\\
    The modeled distribution of the precessing trajectory for the superluminal
    components of jet-A is shown in Figure 3 (left panel) for precession
     phases $\phi$=0.0, 1.0, 2.0, 3.0, 4.0, 5.0\,rad, respectively. The 
    projected trajectories of B5, K09 and K14 within the jet-cone
    are shown in the right panel.\\
    As shown in the previous paper (Qian et al. \cite{Qi21}), five superluminal 
    components (B4, B5, K2, K3, K09 and K14) of jet-A were found to undergo
     accelerated/decelerated motion, revealing  increase/decrease 
     in their bulk Lorentz factor and Doppler factor. However,  
     their flux evolution due to the Doppler-boosting effect has not been
     investigated. Here we shall model-fit the 43\,GHz light curve for knot B4
    in terms of its Doppler-boosting effect in combination with the explanation 
    of its kinematics.
    \begin{figure*}
  \centering
  \includegraphics[width=5.6cm,angle=-90]{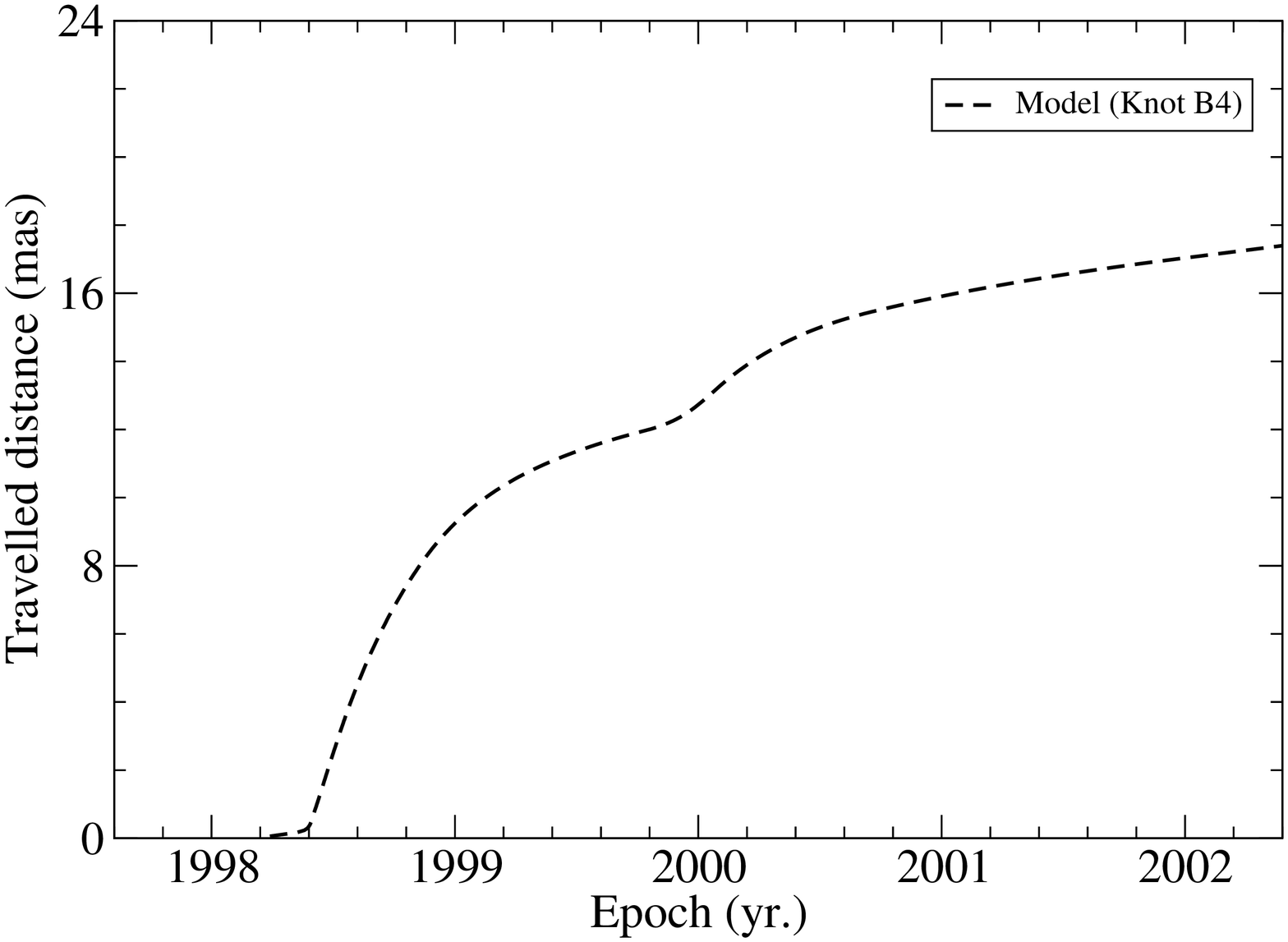}
  \includegraphics[width=5.6cm,angle=-90]{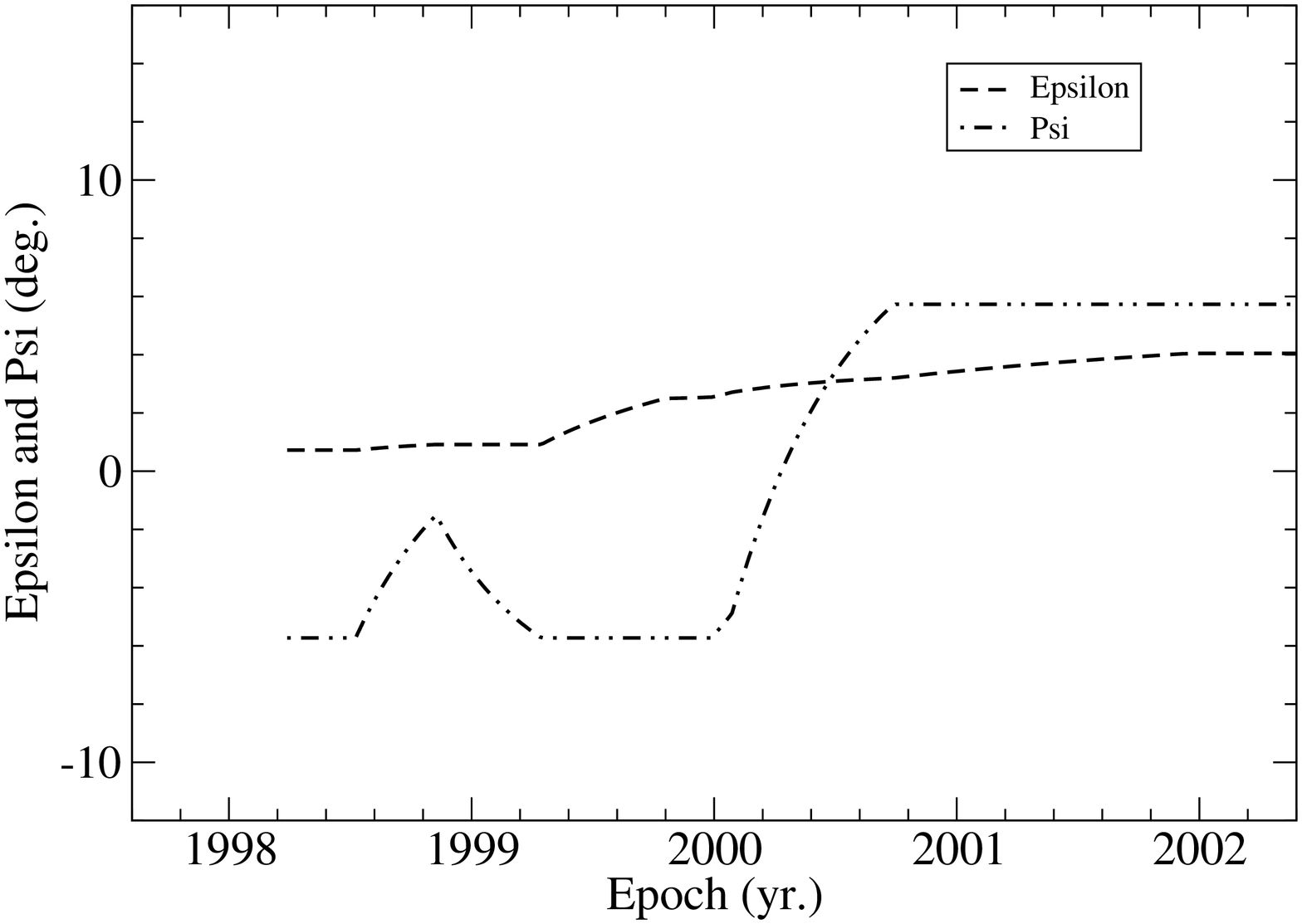}
  \caption{Knot B4: the modeled traveled distance Z(t) along the jet axis 
   (left panel) and the modeled curves for the parameter 
   $\epsilon(t)$ and $\psi(t)$ defining the plane where the jet-axis locates. 
   Before 1998.52 $\epsilon$=$0.72^{\circ}$ and $\psi$=--$5.73^{\circ}$, 
   B4 moved along the precessing common trajectory. After 1998.52 B4 started 
   to move along it own individual track. During the two radio bursts 
   ($\sim$1998.4--1999.4 and $\sim$1999.9--2001.2) parameter $\psi$ varied 
   quite largely.}
  \end{figure*}
  \begin{figure*}
  \centering
  \includegraphics[width=7cm,angle=-90]{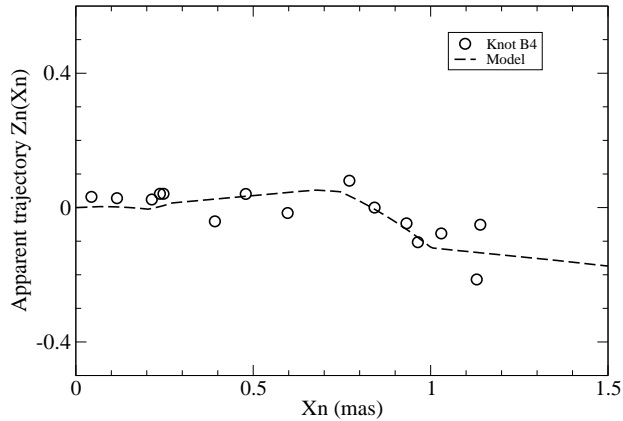}
  \caption{Knot B4: the model-fit of the observed trajectory $Z_n(X_n)$. Within
  $\sim$0.07\,mas of the coordinate $X_n$ B4 moved along the precessing common 
  trajectory, while beyond that it moved along its own individual track,
     deviating from the common track. A prominent curvature occurred at 
    $r_n\simeq$0.7--1.2\,mas.} 
  \end{figure*}
  \begin{figure*}
  \centering
  \includegraphics[width=5.6cm,angle=-90]{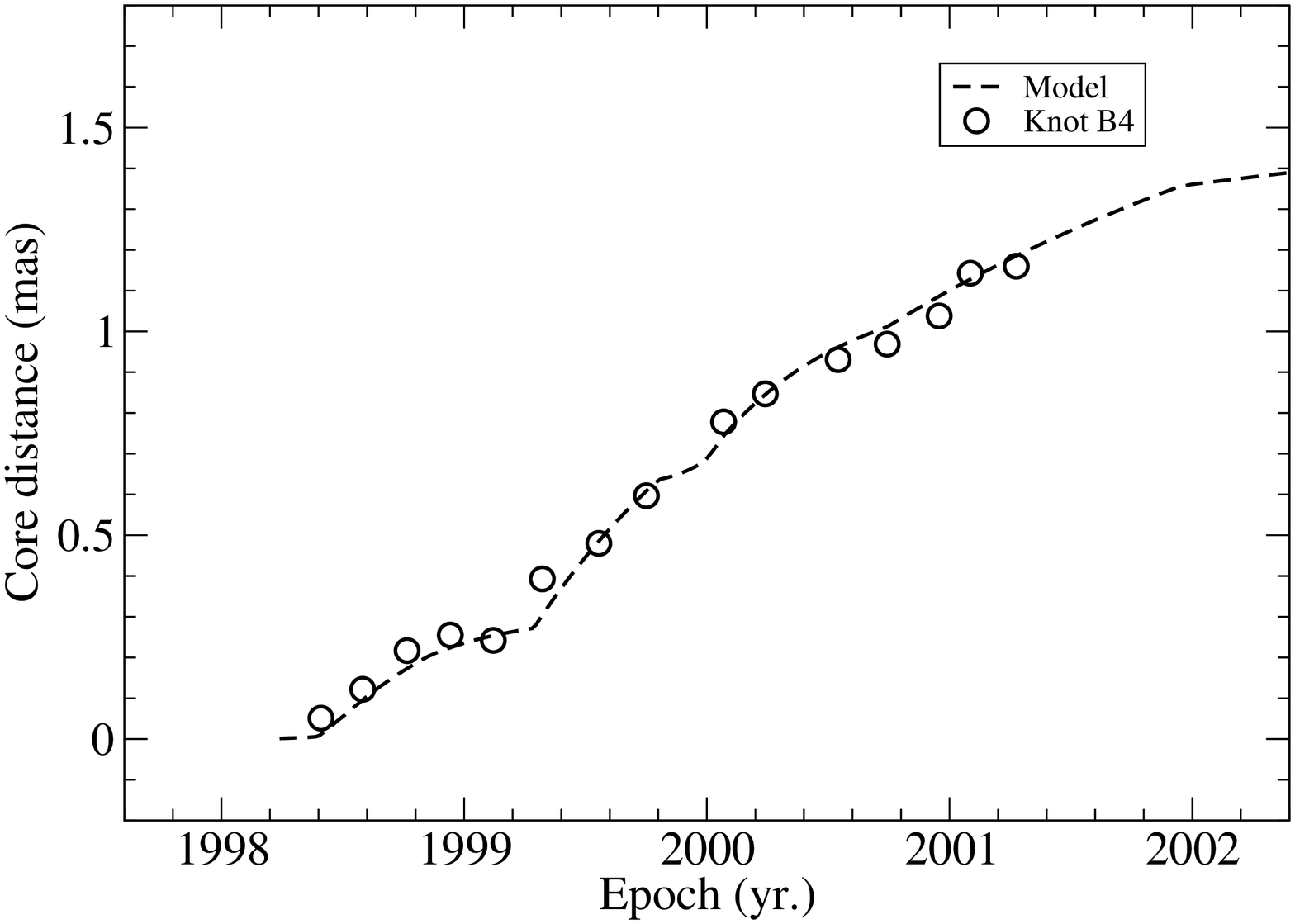}
  \includegraphics[width=5.6cm,angle=-90]{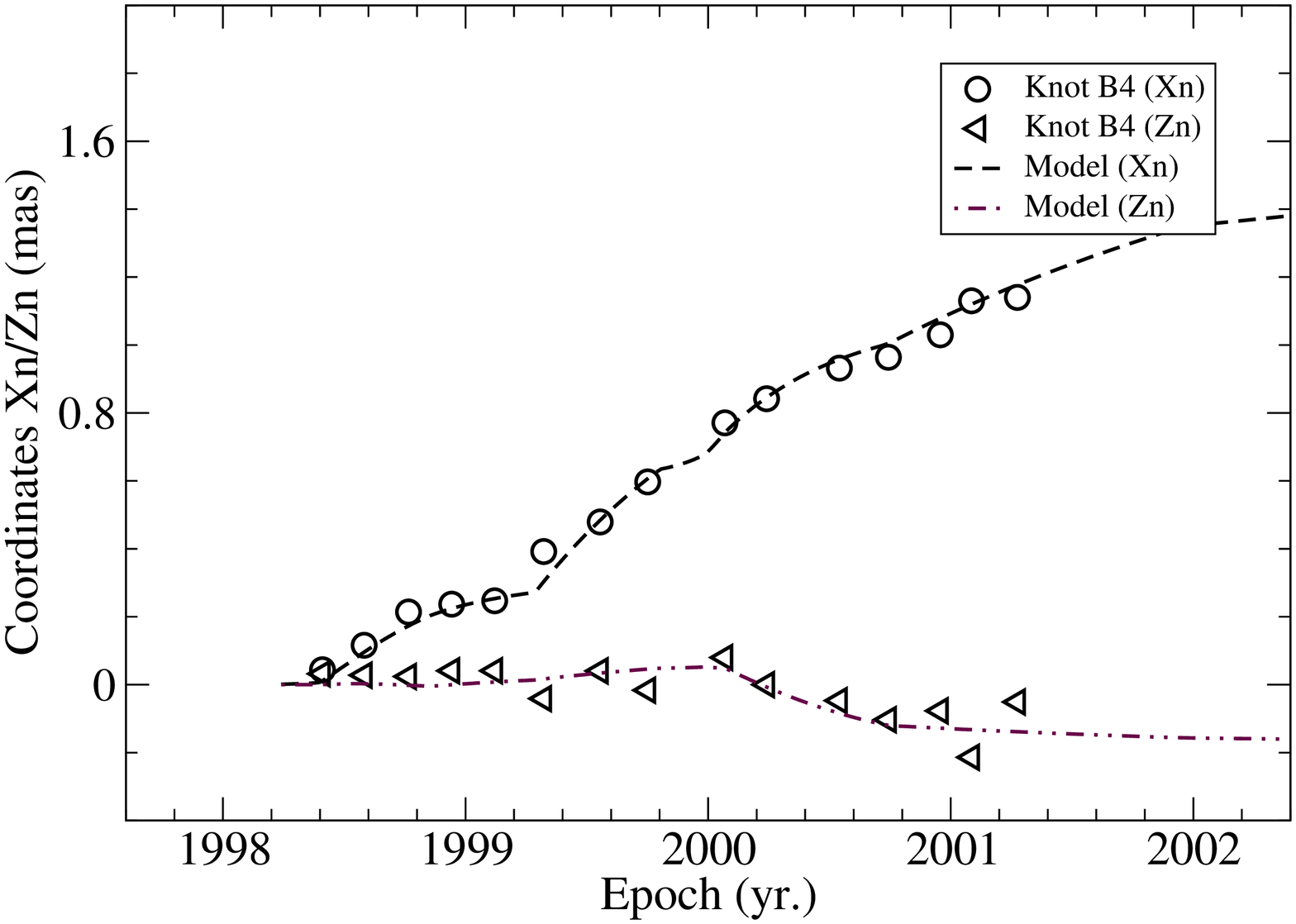}
  \caption{Knot B4: the model-fits of the core separation $r_n(t)$ and 
   coordinates $X_n(t)$ and $Z_n(t)$. These fits are almost perfect (especially
   the fit for $Z_n(t)$) and much better than those previously presented in
    Qian et al. (\cite{Qi21}) due to carefully having taken the details in 
   its kinematic behavior into full account (especially details in
    $r_n(t)$, d$r_n$/dt and $Z_n(t)$). }
   \end{figure*}
  \begin{figure*}
  \centering
  \includegraphics[width=5.6cm,angle=-90]{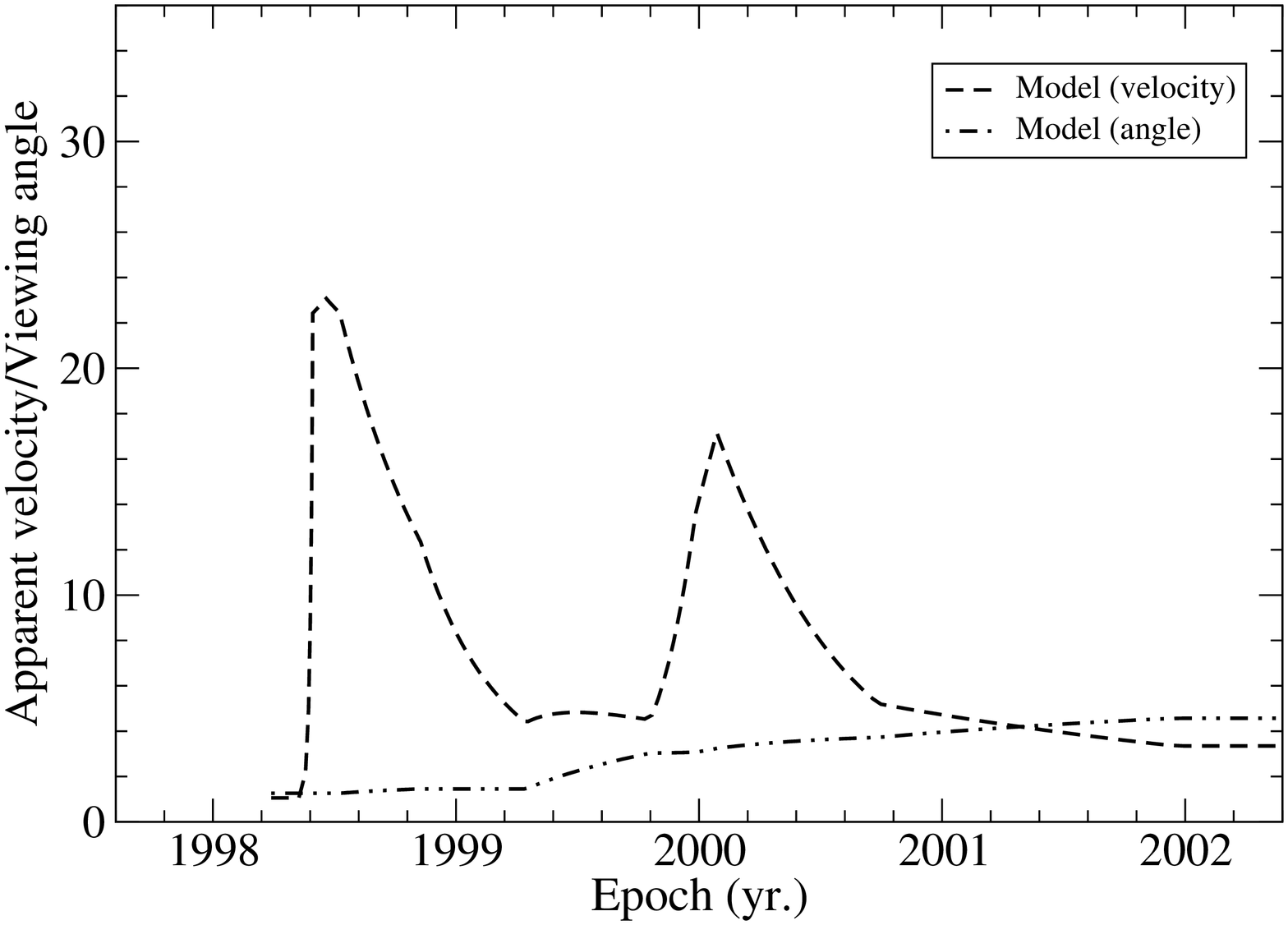}
  \includegraphics[width=5.6cm,angle=-90]{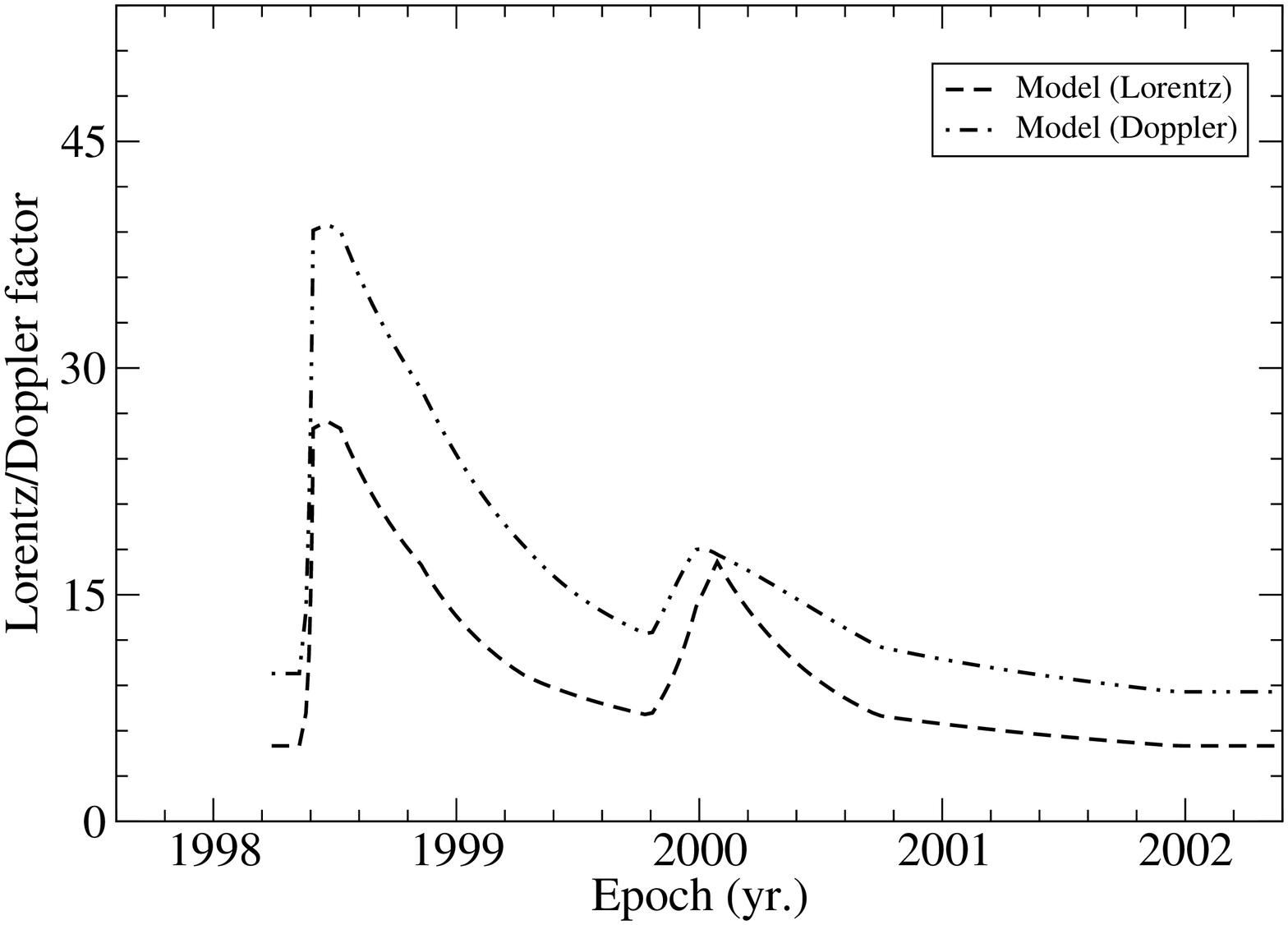}
  \caption{Knot B4. Left panel: the model-derived apparent velocity 
    $\beta_{app}(t)$ and viewing angle $\theta(t)$. Right panel: the 
   model-derived bulk Lorentz factor $\Gamma(t)$ and Doppler 
   factor $\delta(t)$. $\Gamma(t)$, $\delta(t)$ and $\beta_{app}(t)$ all 
   show a double-peak structure closely related to its two radio bursts. 
   At the first peak (epoch 1998.5)
   $\delta_{max}$=39.5, $\Gamma_{max}$=26.5, $\beta_{app,max}$=23.1 and
   $\theta$=$1.26^{\circ}$, while at the second peak (epoch 2000.0) 
    $\delta$=$\delta_{max}$=18.0, $\Gamma$=14.9, $\beta_{app}$=14.5 and
      $\theta$=$3.1^{\circ}$. The maximum Lorentz 
    factor was not coincident with the maximum Doppler factor: at 2000.1
    $\Gamma$=$\Gamma_{max}$=17.2, $\delta$=17.7, $\beta$=$\beta_{app,max}$=17.1
    and $\theta$=$3.24^{\circ}$. Our model-fitting results for the apparent
    velocity are quite different from those given in Jorstad et al. 
   (\cite{Jo05}; cf. its Fig.30) which were determined by using  polynominal
    approximations. Figs.5--7 demonstrate that the entire kinematics observed 
    within core separation ${r_n}{\simeq}$1.2\,mas (corresponding to the 
    traveled distance Z=16.3\,mas$\simeq$125\,pc) can be well explained in
    terms of our precessing nozzle scenario.}
   \end{figure*}
  \subsection{Knot B4: model-fitting results of kinematics} 
    The model-fitting results of its kinematics are shown in Figures 4-7.\\
    The traveled distance Z(t) along the jet-axis (left panel) and the
    modeled curves for parameters $\epsilon(t)$ and $\psi(t)$ are shown in 
    Fig.4. Before 1998.52, or core distance $r_n{\leq}$0.066\,mas
    (Z$\leq$23.0\,pc), $\epsilon$=$0.72^{\circ}$ and $\psi$=--$5.73^{\circ}$,
    knot B4 moved along the precessing common trajectory, while after 1998.52
     it started to move along its own individual trajectory, where 
    parameter $\psi$ showed quite large changes during the two radio bursts in
    (1998.4--1999.4) and (1999.9--2001.2).\\  
  In the model fitting its precession phase was assumed to be 
   $\phi_0$=4.58\,rad and the corresponding ejection time $t_0$=1998.24, 
   well consistent with the ejection
   time 1998.36$\pm$0.07 derived by Jorstad et al. (\cite{Jo05}).\\
  It can be seen from Figs. 5 and 6 that its entire kinematic features
   (including trajectory $Z_n(X_n)$, core separation $r_n(t)$, coordinates 
  $X_n(t)$ and $Z_n(t)$) are all well fitted extending to core separation 
    $r_n{\sim}$1.2\,mas. The model-derived apparent velocity $\beta_{app}(t)$,
     viewing angle $\theta(t)$, bulk Lorentz factor $\Gamma(t)$ and Doppler
    factor $\delta(t)$ are shown in  Fig.7. $\Gamma(t)$, $\delta(t)$ and 
    $\beta_{app}(t)$ all have a double-peak structure, corresponding to the 
   double-peak structure of the light curve (Fig.8).
   The viewing angle $\theta(t)$ varied from ${\sim}1.3^{\circ}$ 
   (1998.3) to ${\sim}4.6^{\circ}$ (2002.0).\\
     \begin{figure*}
     \centering
     \includegraphics[width=5.6cm,angle=-90]{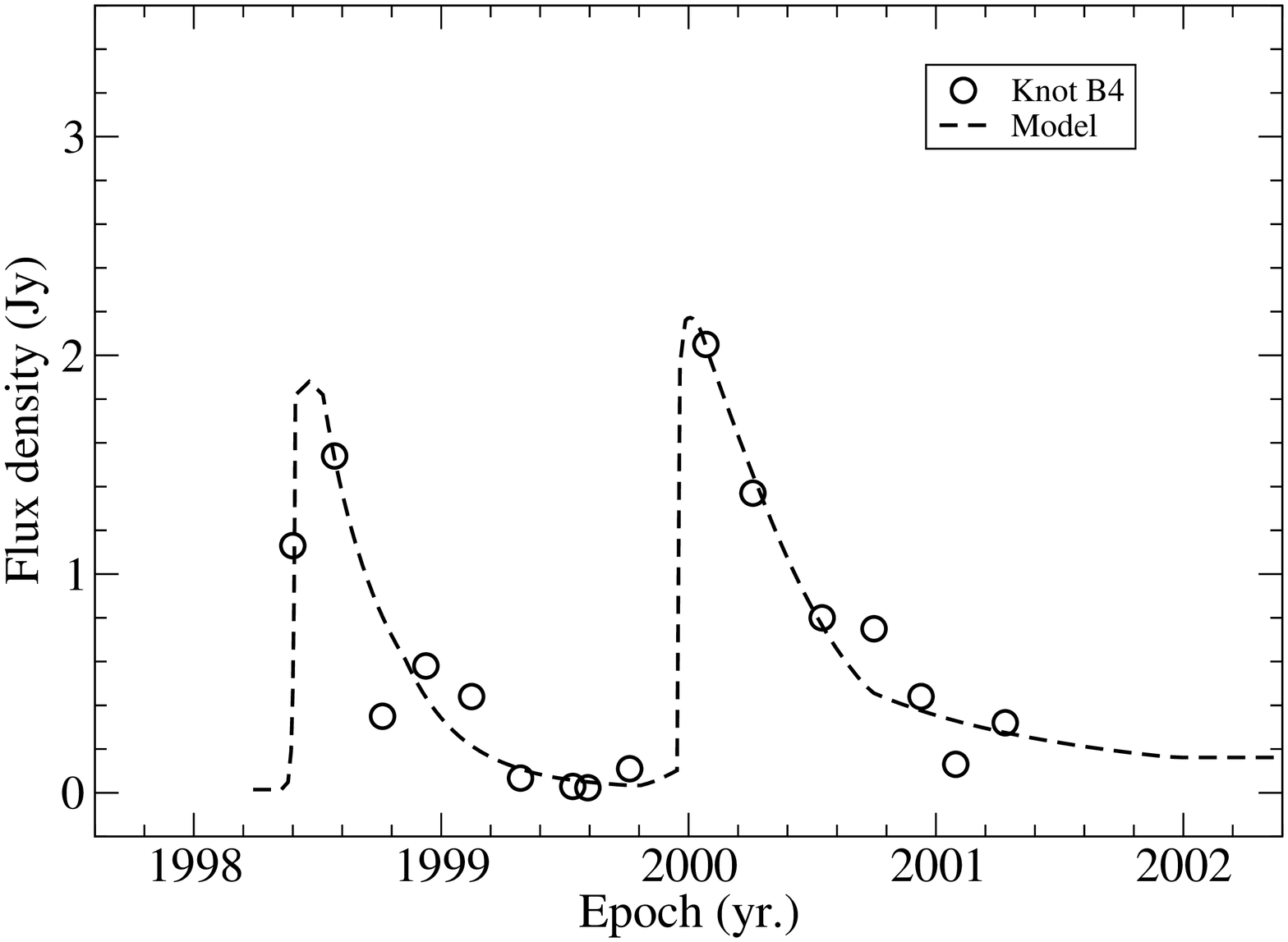}
     \includegraphics[width=5.6cm,angle=-90]{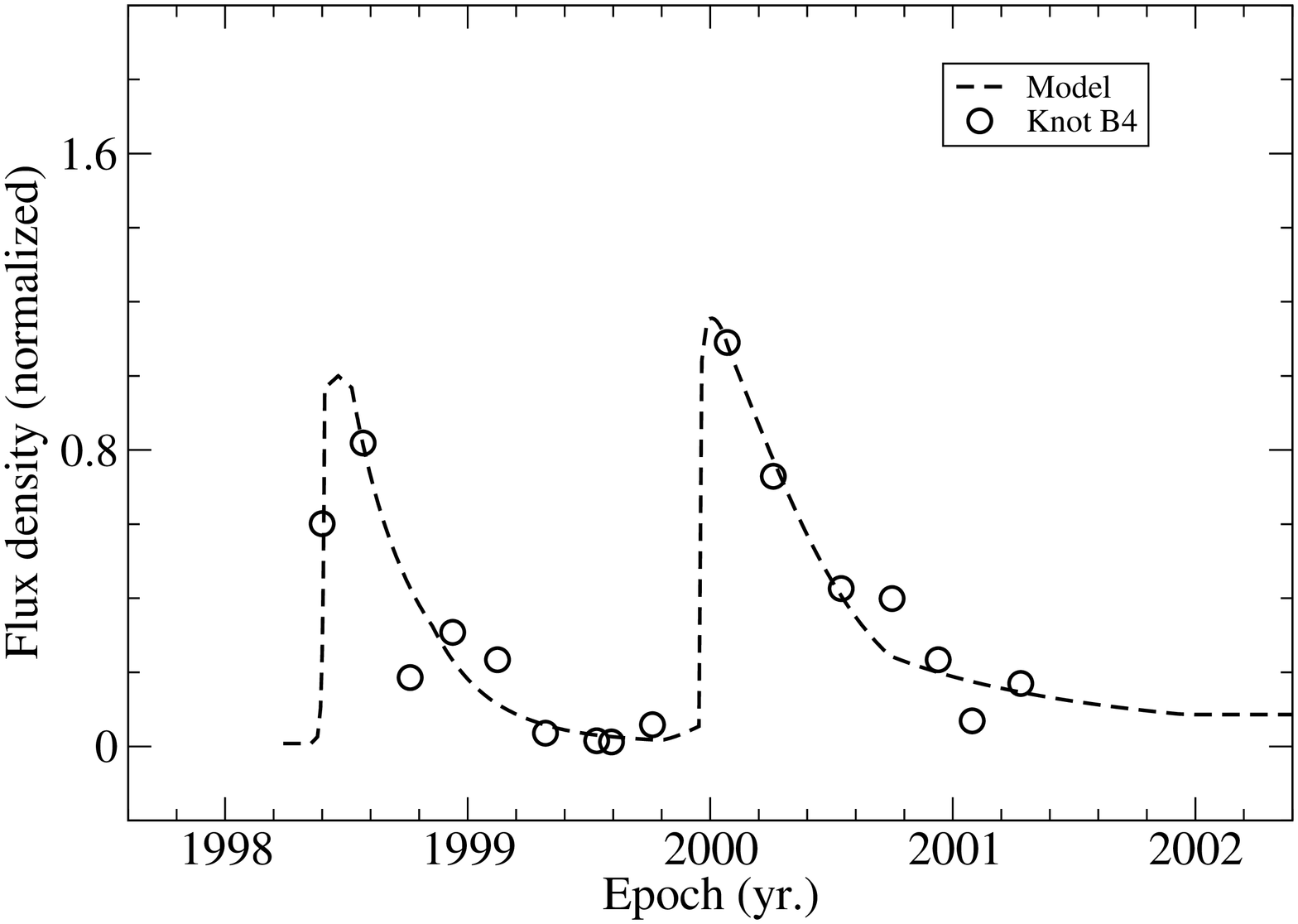}
     \caption{Knot B4. Left panel: the model fit of the 43\,GHz light curve.
      The modeled peak flux densities are 1.88\,Jy (at 1998.5) and 2.17\,Jy 
    (at 2000.0), respectively. The intrinsic flux density of the first burst
      was 4.86$\mu$Jy.  Right panel: the light curve normalized
      by the modeled peak flux density of the first burst 
     is well fitted by the Doppler boosting profile 
     $[\delta(t)/\delta_{max}]^{3+\alpha}$
     ($\alpha$ was asssumed to be 0.5). The second burst had its Doppler factor
     ($\delta_{max}$=18.0) much smaller than the first burst
     ($\delta_{max}$=39.5), while its observed flux density (2.17\,Jy) was 
     larger than that (1.88\,Jy) of the first burst. Thus in the model 
     fitting of the flux evolution the intrinsic flux density of the second
      burst was assumed to be eighteen times that of the first burst 
     (i.e., 87.5$\mu$Jy for the second burst).}
     \end{figure*}
    \subsection{Knot B4: Flux evolution and Doppler-boosting effect}
     B4 produced two radio bursts during 1998.4--1999.4 and 
     1999.9--2001.2. In order to interpret the entire lightcurve, the
     model parameters ($\epsilon(t)$, $\psi(t)$ and $\Gamma(t)$) were carefully
     and consistently selected and we paid  much more attention on the
     details in the observed kinematic behavior (especially on the details
     in $r_n(t)$, $dr_n/dt$ and $Z_n(t)$).\\
       The model-fit results of the entire light curve (including both the 
     radio bursts) are shown in Figure 8.
       For the first burst, the  epoch of the modeled peak $t_{max}$=1998.47
     with the  maximum Doppler factor $\delta_{max}$=39.49 and the maximum 
    flux density $S_{max}$=1.88\,Jy, while its intrinsic flux density 
    $S_{int}$=4.86\,$\mu$Jy.\\
       For the second burst, the  epoch of the modeled peak $t_{max}$=2000.01
     with the maximum Doppler factor $\delta_{max}$=18.02 and maximum flux
      density $S_{max}$=2.17 Jy, while its intrinsic flux density
     $S_{int}$=87.5\,$\mu$Jy. It should be noted that the second burst 
    originated from its re-acceleration in the  convergence/collimation 
    region near the position [$r_n$=0.6--1.2\,mas, PA=$\sim$${-90^{\circ}}$],
    where parameter $\psi$  rapidly increased, resulting in the southwest
    curvature of its trajectory (Figs.\,4--6).
    \section{Knot B6: Model fitting of kinematic behavior and 43\,GHz light 
    curve}
    \begin{figure*}
    \centering
    \includegraphics[width=5.6cm,angle=-90]{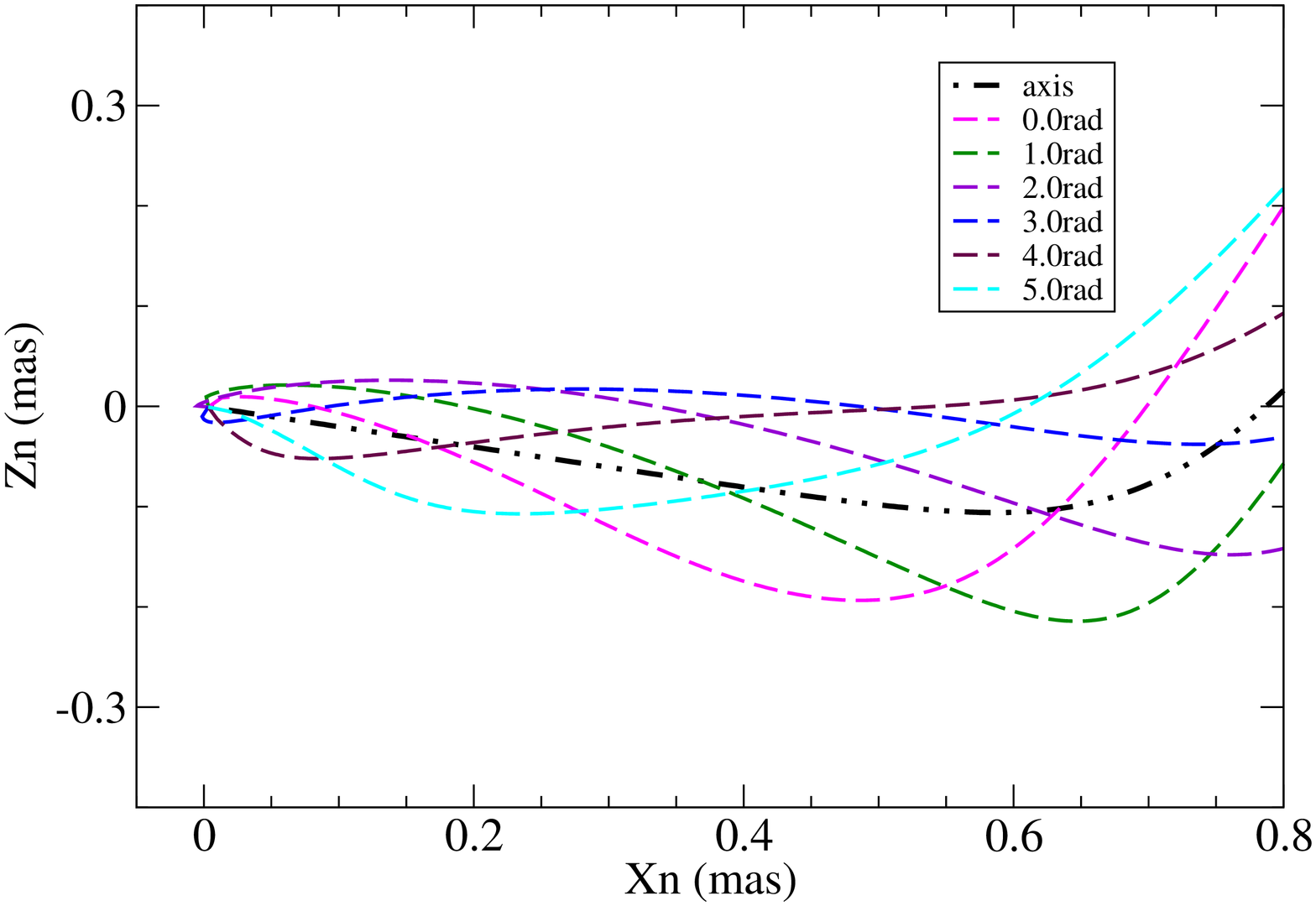}
    \includegraphics[width=5.6cm,angle=-90]{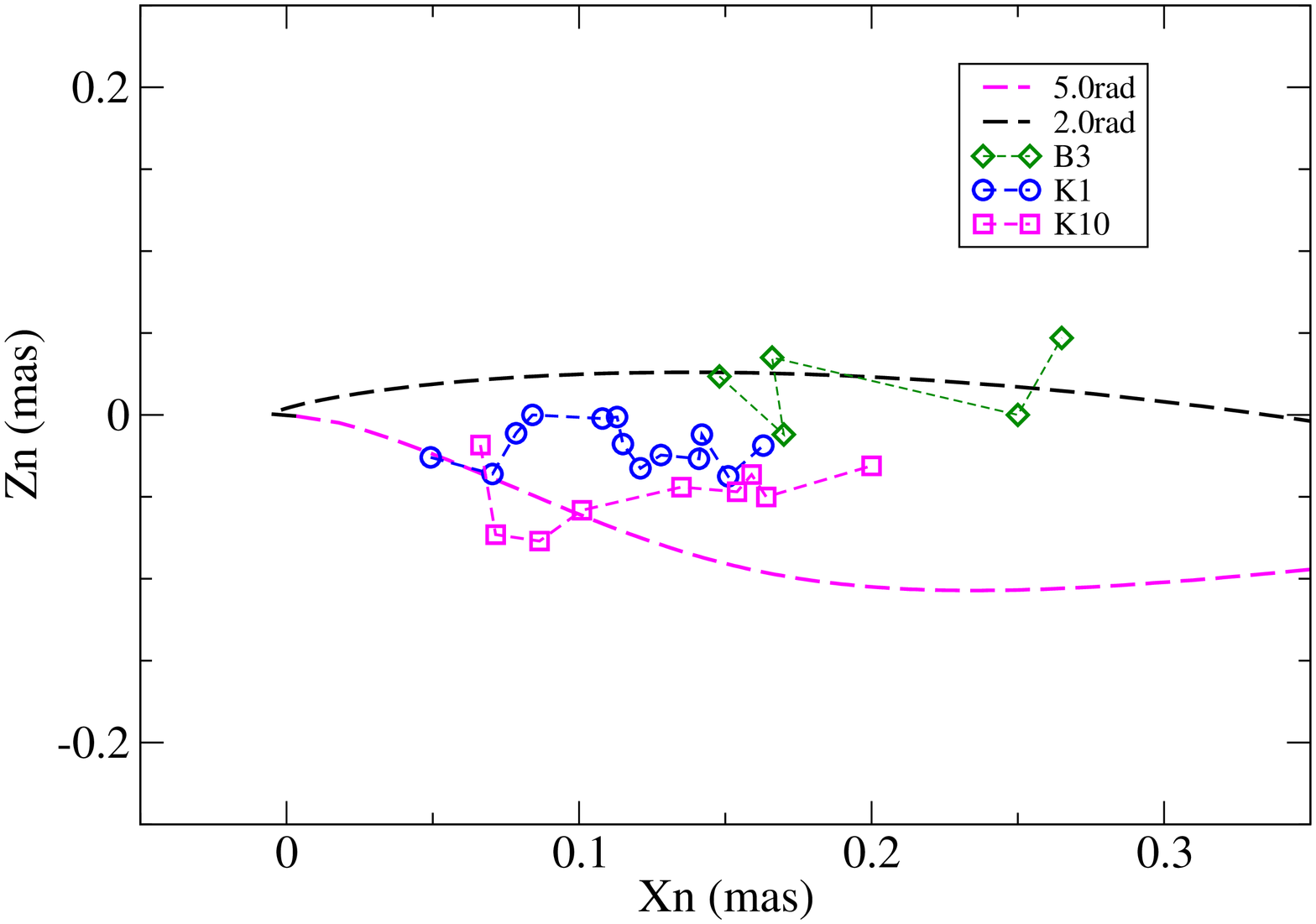}
    \caption{Jet-B. Left panel: the modeled distribution of the precessing 
    common trajectories for the superluminal components at precession phases
    $\phi$=0.0, 1.0, 2.0, 3.0, 4.0 and 5.0\,rad, repectively. Right panel:
    the projected trajectories of knots B3, K1 and K10 within the jet cone. The
    jet-axis is at position angle --$101.5^{\circ}$.}
    \end{figure*}
     As in the previous work superluminal 
   knots B6 and B1, B2, B3, K1, K10, K16 were assumed to be
    ejected by the nozzle of jet-B. For jet-B the modeled distribution of the
   precessing common trajectories for its superluminal components is shown in 
   Fig.9. We shall use the same model parameters as before (Qian et al. 
   \cite{Qi21}).\\
   (a) The jet-axis locates in a plane defined by the parameters 
   $\epsilon$=0.0126\,rad=$0.72^{\circ}$ and $\psi$=0.20\,rad=$11.46^{\circ}$.\\
   (b) The shape of the jet-axis is defined by a set of parameters: $\zeta$=2,
    $p_1$=0, $p_2$=1.34$\times$$10^{-4}$/mas, $z_t$=66\,mas and $z_m$=6.0\,mas
    (cf. equations 1 and 2).\\
   (c) The common precessing trajectory pattern is defined by the
     amplitude parameter A=${A_0}{[\sin(\pi{z_0}/{Z_1})]^{1/2}}$ with 
    ($A_0$=0.182\,mas and $Z_1$=396\,mas) and helical phase
    $\phi$=$\phi_0$--$(z_0/Z_2)^{1/2}$ with $Z_2$=3.58\,mas and $\phi_0$ being
     the precession phase which is related to its ejection time $t_0$
    as follows:
      \begin{equation}
        {\phi_0}(rad)=0.42+\frac{2\pi}{T_0}(t_0-1994.46)
      \end{equation}
    As shown in the previous paper (Qian et al. \cite{Qi21}), four superluminal
    components (B2, B6, K1 and K16) of jet-B were found to participate in
   accelerated/decelerated motions, showing the trends of increasing/decreasing    in their bulk Lorentz factor and Doppler factor. However, untill now 
    their flux evolution due to the Doppler-boosting effect has not been 
    investigated. Here we  shall model-fit the 43\,GHz light curve for knot B6
    in combination with the explanation of its kinematics.
    \section{Knot B6: model-fitting results of kinematics}
   The model fitting results of its kinematics are shown in Figures 10--13.\\
   Its ejection time $t_0$=1999.61, well consistent with that
   ($t_0$=1999.80$\pm$0.37) given by Jorstad et al. (\cite{Jo05}) and the 
   corresponding precession phase $\phi_0$=3.50\,rad.\\
    The modeled traveled distance Z(t) along the jet-axis and the model-derived
    curves for the parameters $\epsilon(t)$ and $\psi(t)$ are shown in Fig.10.
    Before 2000.54 (or $r_n{\leq}$0.18\,mas) $\epsilon$=$0.72^{\circ}$ and
    $\psi$=$11.5^{\circ}$, knot B6 moved along the precessing common trajectory.
    After 2000.54 $\psi$ quickly decreased to $-4.6^{\circ}$ and B6 started to
    move along its own individual track, deviating from the precessing common 
    trajectory.\\
     The model-fitting results of the observed trajectory $Z_n(X_n)$, distance
    from the core $r_n(t)$, coordinates $X_n(t)$ and $Z_n(t)$ are shown in
    Figs.\,11 and 12. All these kinematic features were well fitted by the
    precessing nozzle scenario.\\
   The model-derived curves for Lorentz factor $\Gamma(t)$, Doppler factor
   $\delta(t)$, apparent velocity $\beta_{app}(t)$ and viewing angle 
   $\theta(t)$ are shwon in Fig.13. $\Gamma(t)$, $\delta(t)$ and 
   $\beta_{app}(t)$ all have a  bump structure. At 2000.30 $\Gamma_{max}$=30.5,
   and $\delta_{max}$=56.0, but $\beta_{app,max}$=21.4 at 2000.6. The viewing
   angle $\theta(t)$ varied in the range 
  [$0.72^{\circ}$(1999.6)--$0.38^{\circ}$(2000.0)--$1.63^{\circ}$(2001.5)].
    \begin{figure*}
    \centering
    \includegraphics[width=5.6cm,angle=-90]{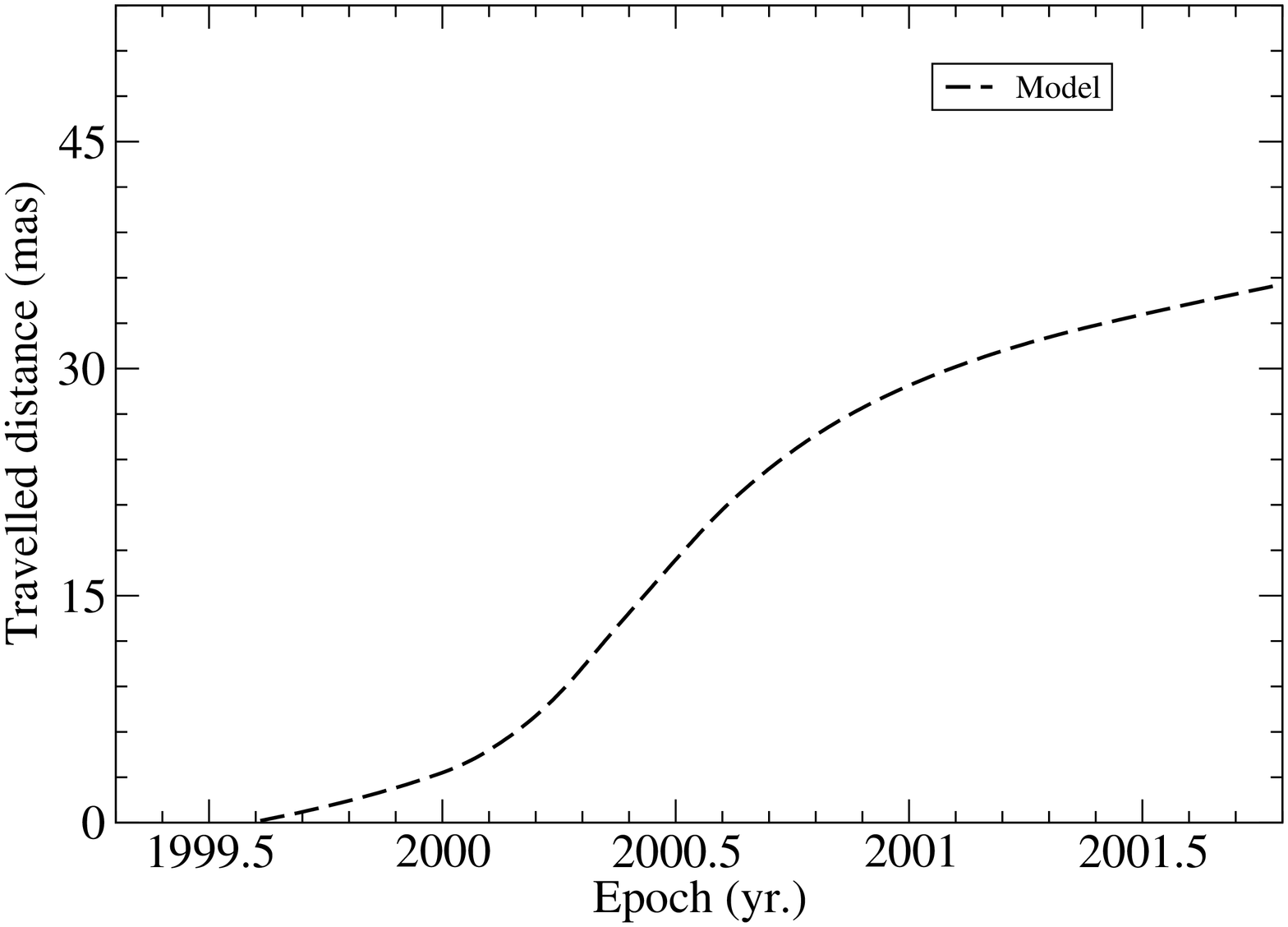}
    \includegraphics[width=5.6cm,angle=-90]{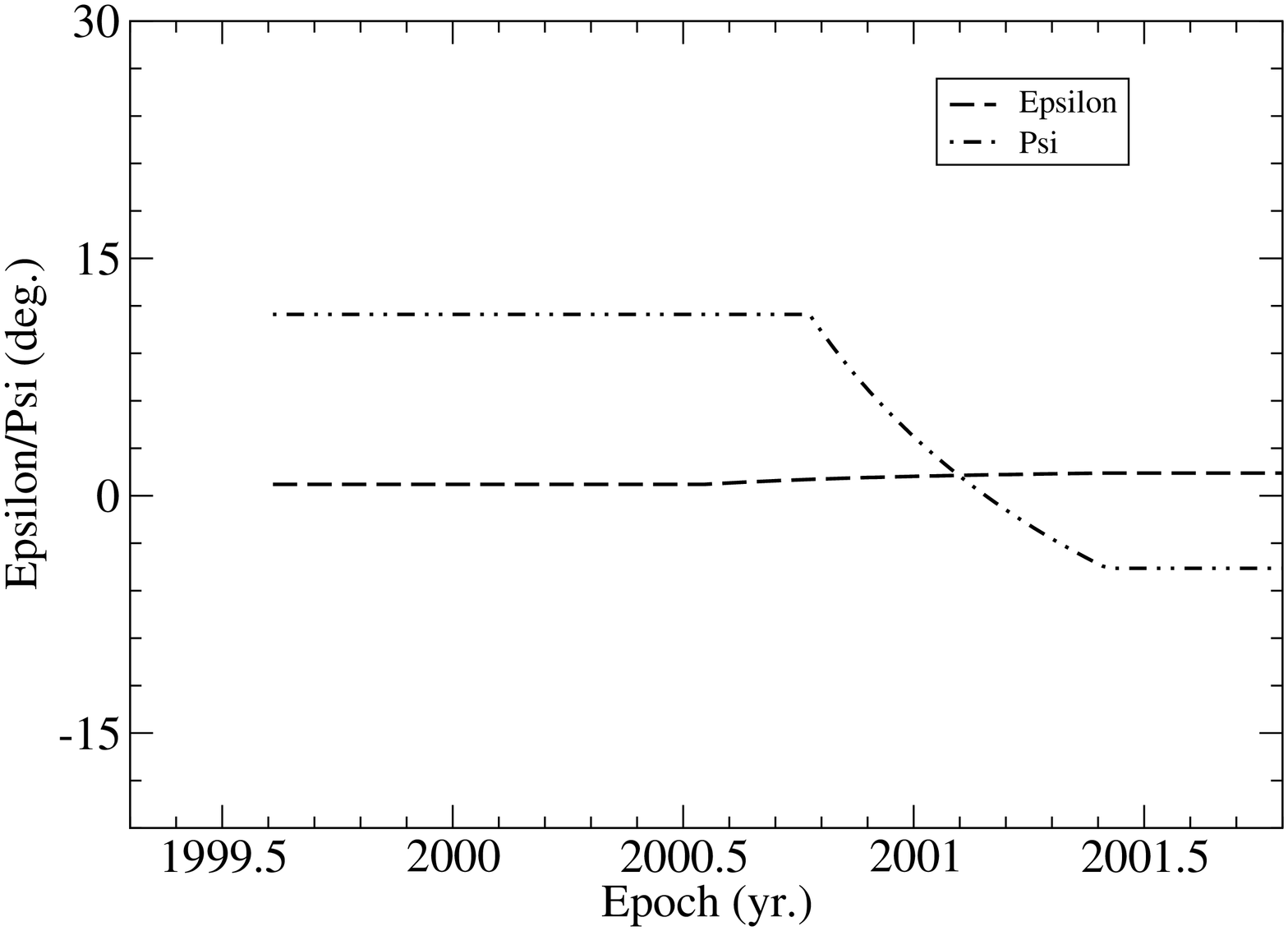}
    \caption{Knot B6: the modeled traveled distance Z(t) along the 
    jet axis (left panel) and the model-derived curves for the 
    parameters $\epsilon(t)$ and $\psi(t)$ which define the plane where the 
    axis of jet-B locates. Before 2000.54 (or $r_n{\leq}$0.18\,mas,
    corresponding to Z$\leq$19.0\,mas=146\,pc) $\epsilon$=$0.72^{\circ}$ and
    $\psi$=$11.5^{\circ}$, knot B6 moved along the precessing common track.
     After 2000.54 $\psi$ rapidly decreased to --$4.6^{\circ}$ (at 2001.4,
     knot B6 started to move along its own individual trajectory, 
    deviating from the precessing common track.}
    \end{figure*}
    \begin{figure*}
    \centering 
    \includegraphics[width=7cm,angle=-90]{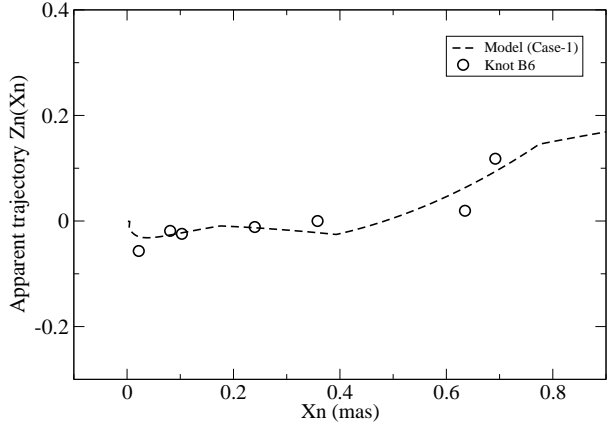}
    \caption{Knot B6: the model fit of the apparent trajectory. Within core
     separation $r_n{\sim}$0.18\,mas (or before 2000.54, corresponding to the 
     traveled distance Z$\leq$19.0\,mas=146\,pc) $\epsilon$=$0.72^{\circ}$ 
     and $\psi$=$11.5^{\circ}$, knot B6 moved along the precessing common 
    trajectory with its precession phase $\phi_0$=3.50\,rad.
    Beyond $r_n$=0.18\,mas it started to move along its own individual track,
    deviating from the precessing common trajectory.}
    \end{figure*}
    \begin{figure*}
    \centering
    \includegraphics[width=5.6cm,angle=-90]{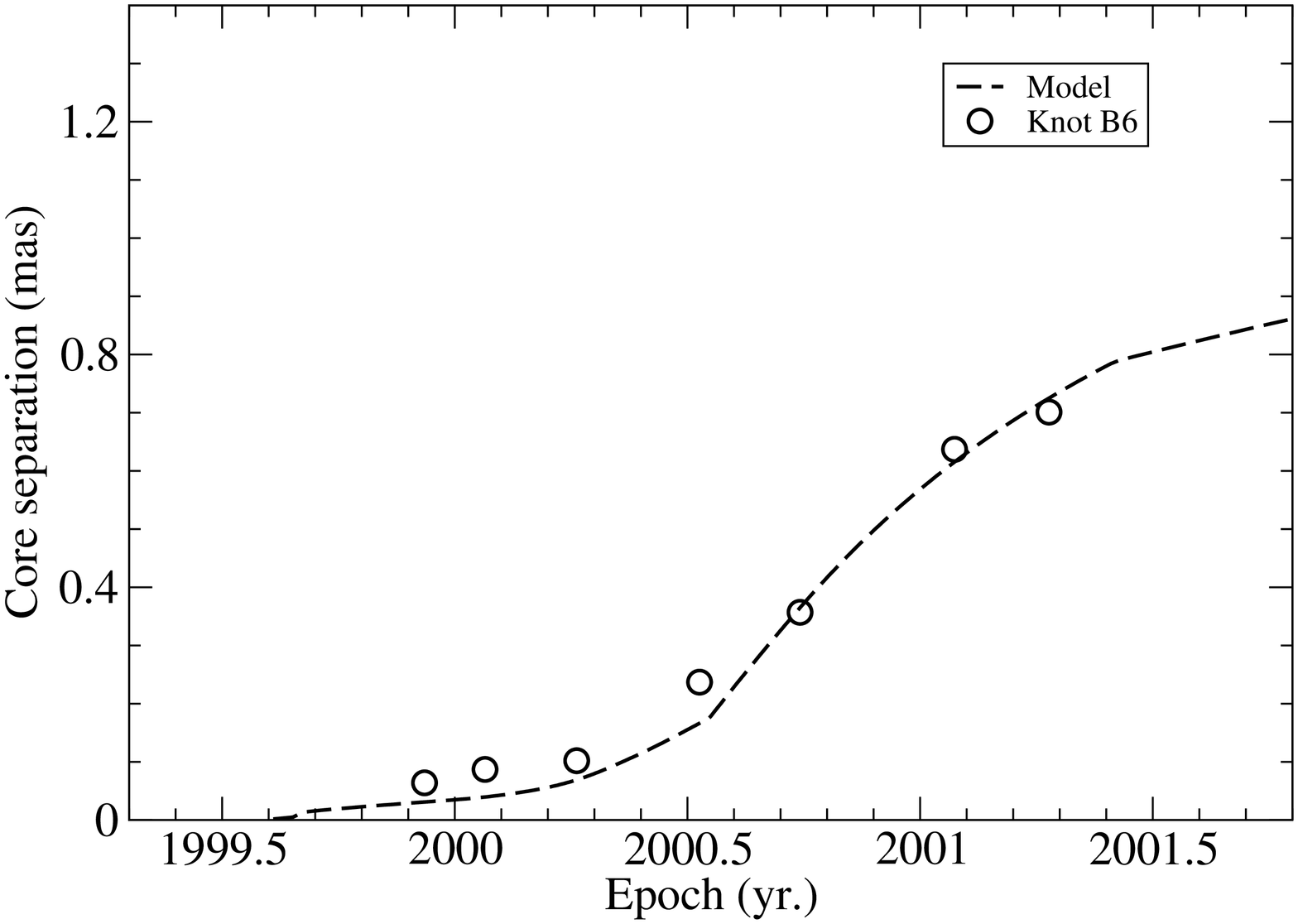}
    \includegraphics[width=5.6cm,angle=-90]{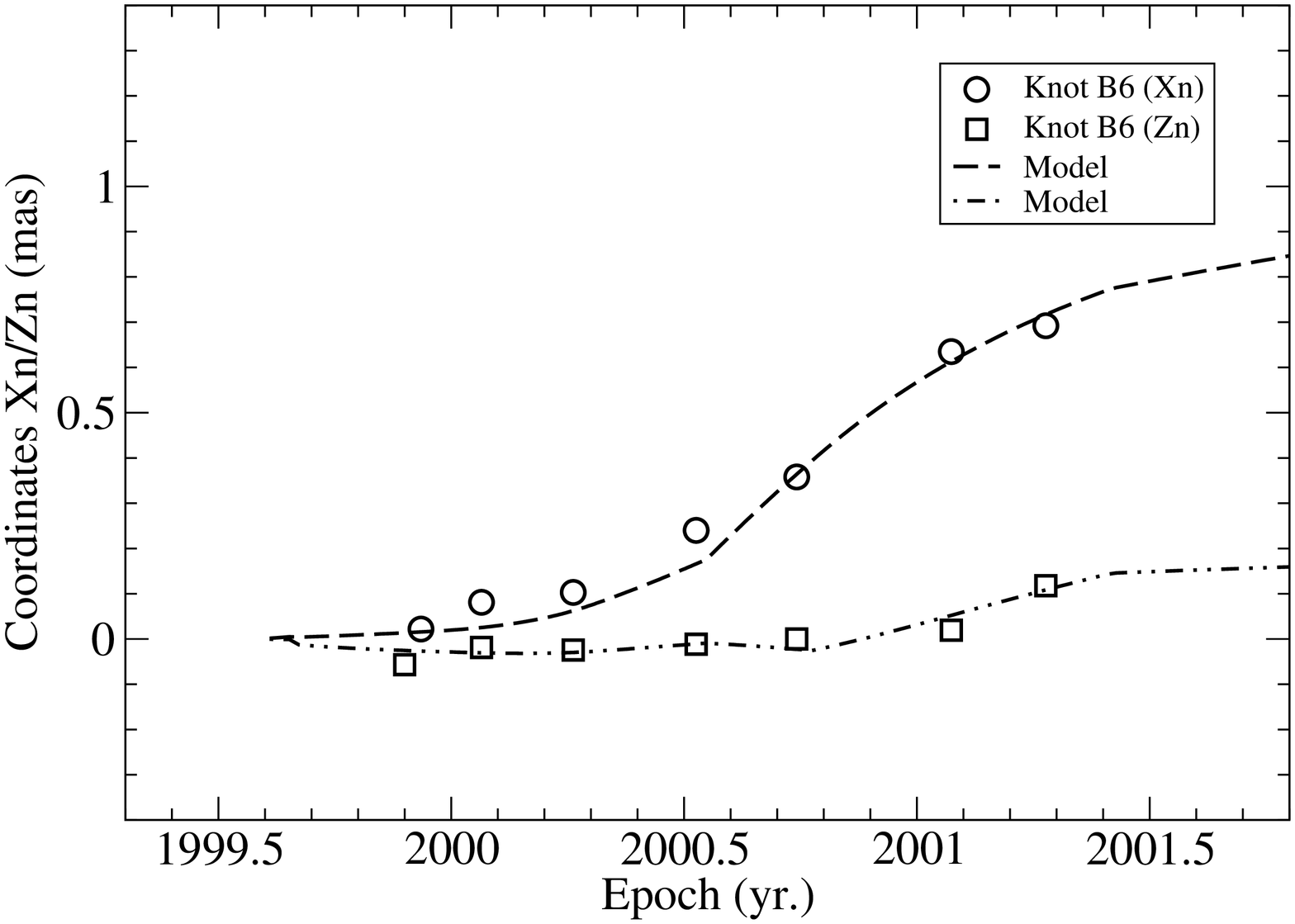}
    \caption{Knot B6: the model-fits of the core separation $r_n(t)$ (left
     panel) and the coordinates $X_n(t)$ and $Z_n(t)$ (right panel). 
    All these kinematic features are well fitted by the precessing nozzle 
    scenario. The radio burst occurred during $\sim$2000--2001 is associated 
    with its accelerated/decelerated motion.}
    \end{figure*}
    \begin{figure*}
    \centering
   \includegraphics[width=5.6cm,angle=-90]{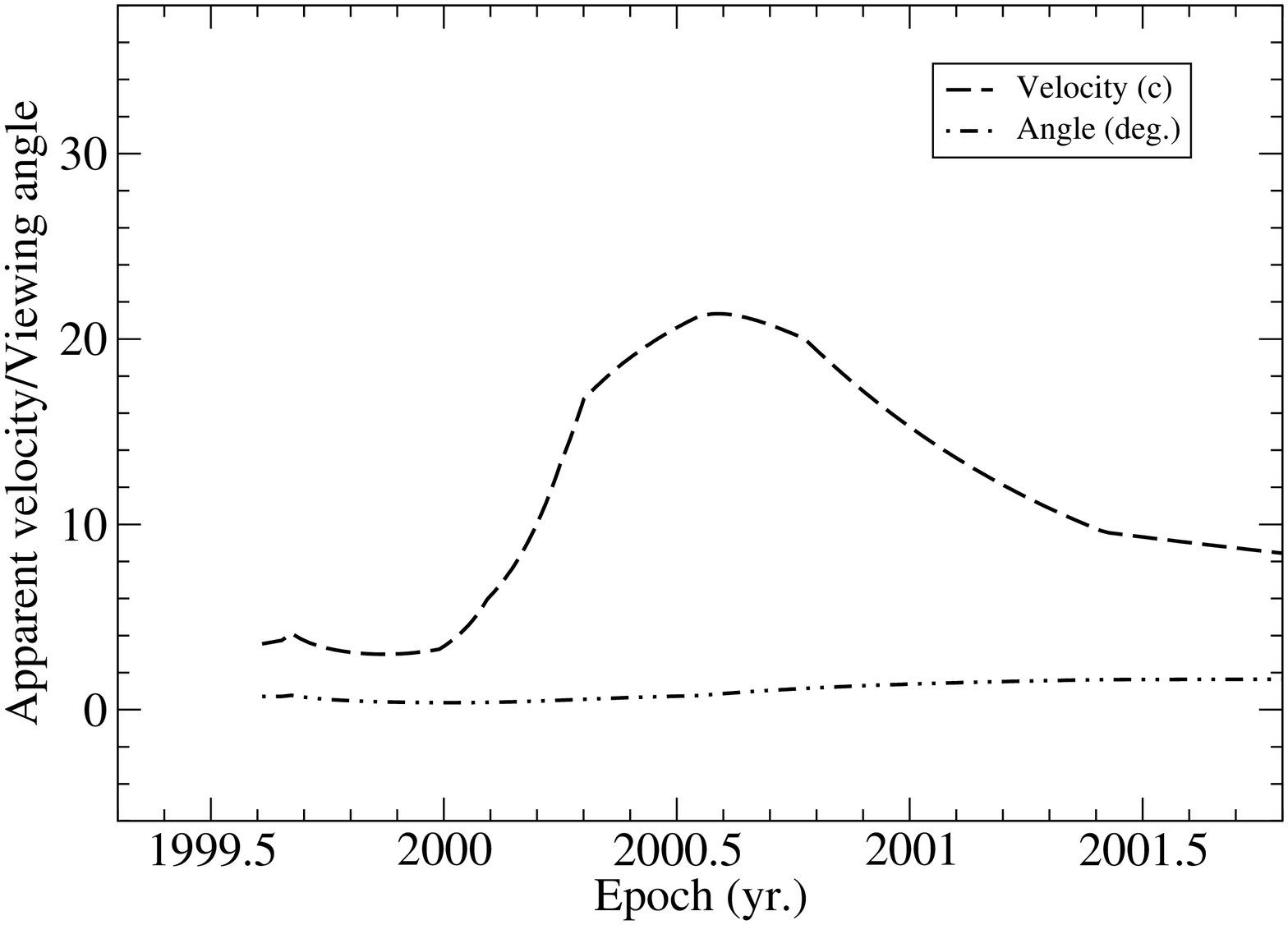}
   \includegraphics[width=5.6cm,angle=-90]{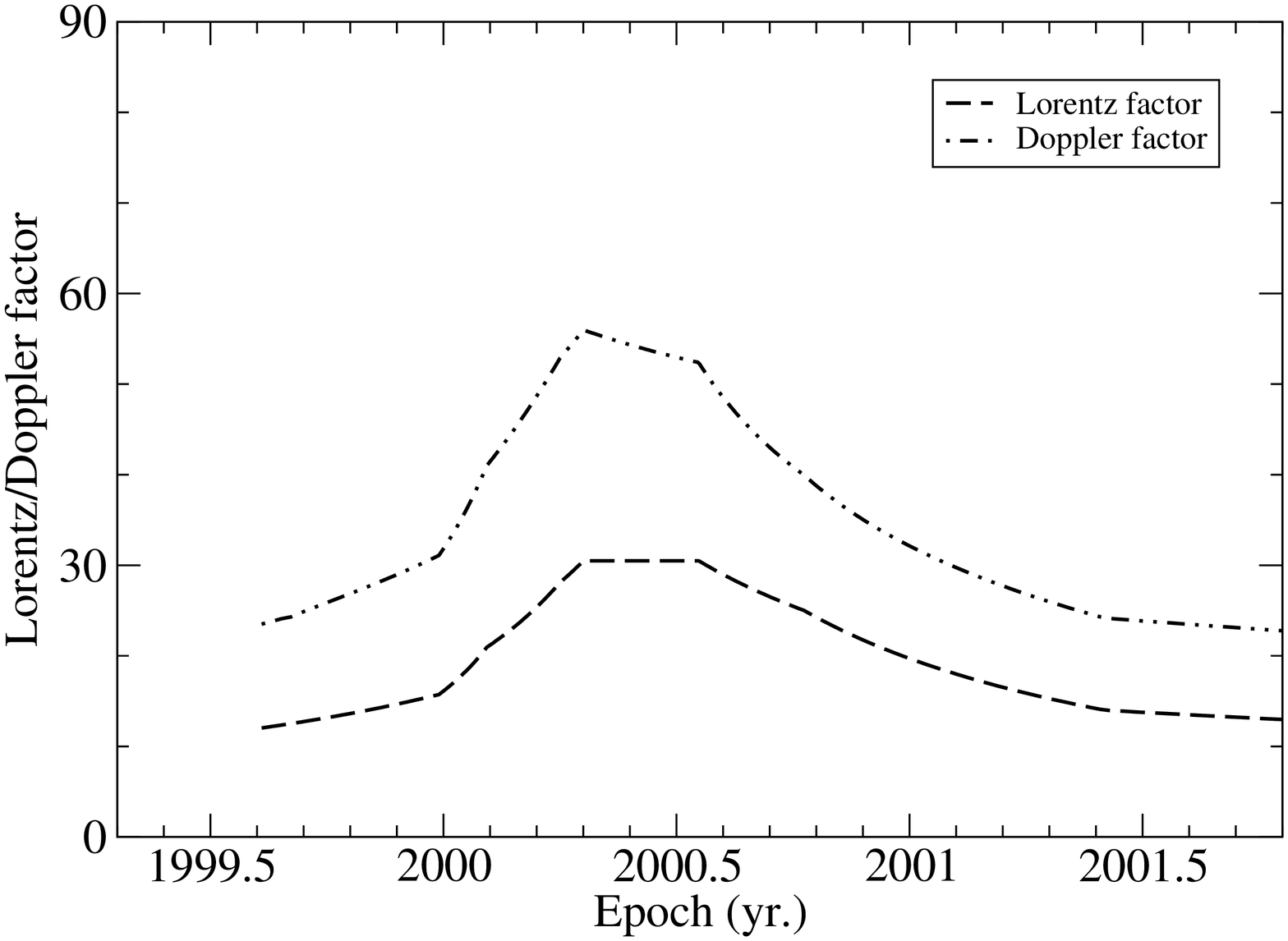}
   \caption{Knot B6: the model-derived apparent velocity $\beta_{app}$ and
    viewing angle $\theta(t)$ (left panel), and the model-derived bulk Lorentz
     factor $\Gamma(t)$ and Doppler factor $\delta(t)$ (right panel).
     $\beta_{app}$, $\Gamma$ and $\delta$ all have a bump structure. At 2000.30
     $\delta$=$\delta_{max}$=55.9 and $\Gamma$=$\Gamma_{max}$=30.5, while
    $\beta$=$\beta_{max}$=21.4 at 2000.59; $\theta(t)$ varied in the range 
    [$0.72^{\circ}$ (1999.6), $0.38^{\circ}$ (2000.0), $1.63^{\circ}$
    (2001.5)]}.
   \end{figure*}
    \begin{figure*}
    \centering
    \includegraphics[width=5.6cm,angle=-90]{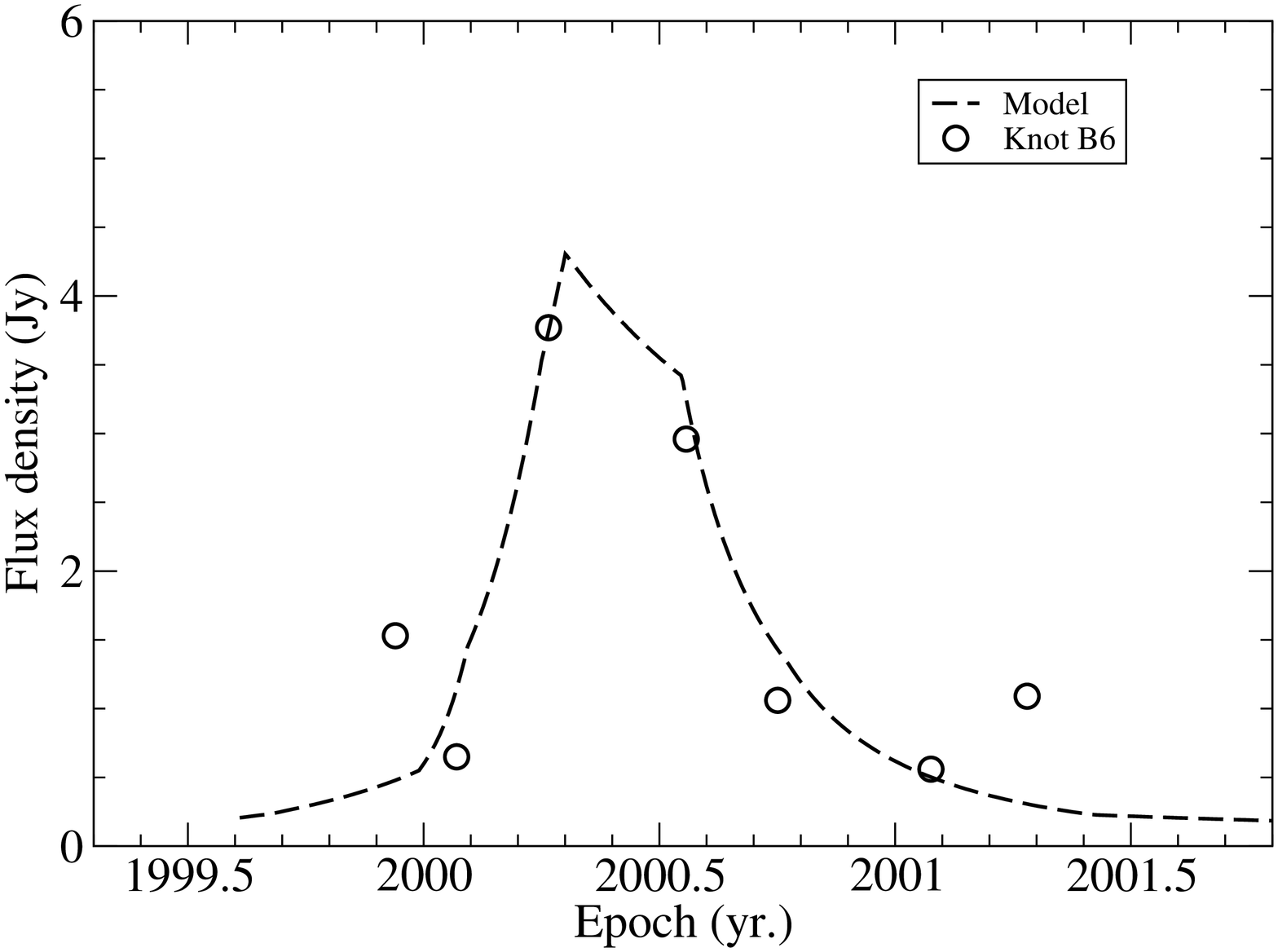}
    \includegraphics[width=5.6cm,angle=-90]{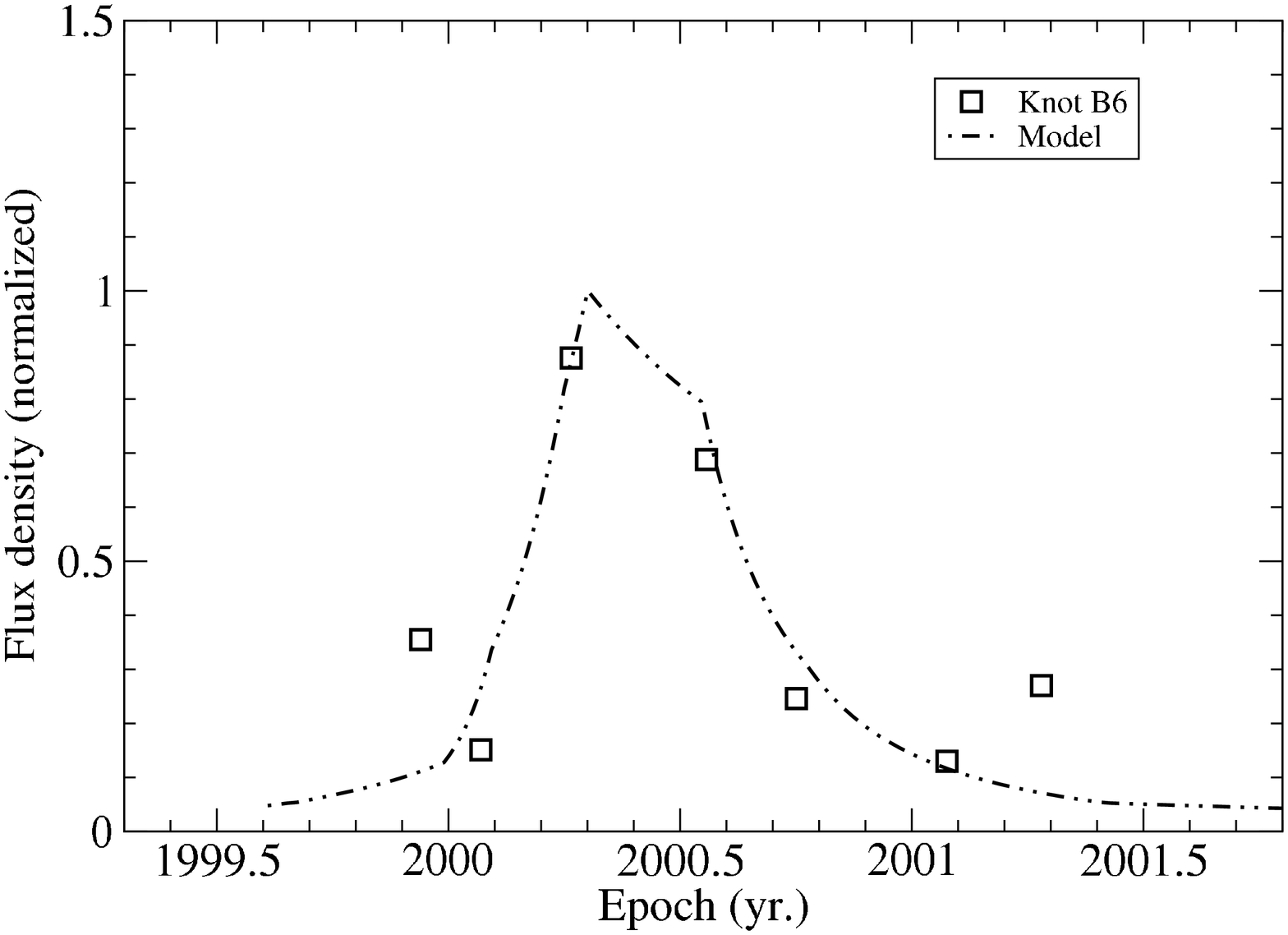}
    \caption{Knot B6: the model-fit of its light curve. The 43\,GHz light curve
    is well fitted by the Doppler-boosting effect with the modeled peak flux 
     density of 4.31\,Jy at 2000.3  and its intrinsic flux density 3.3$\mu$Jy
    (left panel). The light curve normalized by the modeled peak flux
    is well fitted by its Doppler-boosting profile 
    $[\delta/\delta_{max}]^{3+\alpha}$ with an assumed value of $\alpha$=0.5.}
    \end{figure*}
   \subsection{Knot B6: Flux evolution and Doppler-boosting effect}
    As for the case of knot B4 we calculated the observed light curve by using
    equation (19) in order to take account of its Doppler boosting effect.
    The model fit of its light curve is shown in Figure 14. It can be seen that
    the 43\,GHz light curve is well fitted with the modeled peak flux density
    4.31\,Jy at 2000.30 and an intrinsic flux density of 3.3$\mu$\,Jy 
    (left panel). The light curve normalized by the modeled peak flux density
    is well fitted by its Doppler-boosting profile 
   $[\delta(t)/\delta_{max}]^{3+\alpha}$ with a presumed value of 
   $\alpha$=0.5 (right panel).\\
   \section{Knot R3: model fitting results of kinematics and light curve
    observed at 15\,GHz}
   Two characteristics should be emphasized: (1) An arc-like structure was
   detected at 15\,GHz; (2) Its light curve comprises two radio bursts. These
   features are very similar to those of knot B4 observed at 43\,GHz.
    See Appendix.
    \section{Discussion and Conclusion}
    We have applied the precessing jet-nozzle scenario proposed previously
   (e.g. Qian et al. \cite{Qi91}, \cite{Qi14}, \cite{Qi21}; Qian \cite{Qi22a},
    \cite{Qi22b})
    to successfully model-fit the kinematic behavior of the 
    superluminal components
    B4 and B6 in blazar 3C454.3 and nicely explain the 43\,GHz light curves in 
    terms of their Doppler-boosting effect. Thus
     their kinematic properties 
    (including the entire trajectory $Z_n(X_n)$, core separation $r_n(t)$ and
     coordinates $X_n(t)$ and $Z_n(t)$) and their flux evolution have been
    completely interpreted as a whole. Their kinemtic parameters (including 
    bulk Lorentz 
    factor $\Gamma(t)$, Doppler factor $\delta(t)$, viewing angle $\theta(t)$
   and apparent velocity $\beta_{app}(t)$ as function of time) have been
    correctly determined.\\
     In order to achieve this goal, we have carefully taken some significant
    details in their kinematic behavior (e.g.,  details in 
  the observed core-separation curve ${r_n(t)}$ and in the coordinate $Z_n(t)$)
  into full account, which were neglected in the previous work (Qian et al. 
   \cite{Qi21}). \\
   Obviously, similar studies can also be done for more  components 
   (e.g., B5, K3, K09, K14 (jet-A) and B2, K1, K16 (jet-B)) to associate their
   flux evolution with their Doppler boosting effect, because the modeled
    curves of Lorentz/Doppler factor for these components were already 
   derived approximately in Qian et al. (\cite{Qi21}). It seems that the 
   optical light curves of superluminal knots observed in 3C454.3 (Jorstad 
   et al.  \cite{Jo10}, \cite{Jo13}) could also be studied
    within the framework of our precessing nozzle scenario, if the precessing
    common trajectory patterns (e.g., helical patterns on much smaller scales)
    are correctly selected (cf. Qian \cite{Qi18b}).\\
    The full explanation of the kinematics and flux evolution observed at 
    43\,GHz for both B4 and B6 in terms of the precessing nozzle sceanrio has 
    further clarified the distinct features in 3C454.3:
    (1) its superluminal components could be separated into two groups which 
     have different kinematic and flaring properties. For example, knot B4 
     moved southwest along a curved trajectory extending to core separation 
    of $\sim$1.2\,mas, passing through a convergence/recollimation region and
    producing a major burst. Moreover, the track of knot B4 could be 
    connected to that of knot D at
    core distance $\sim$6\,mas (cf. Fig.1).  In contrast, knot B6 moved 
    northwest along a track only extending
    to core distance of $\sim$0.7\,mas without any flaring activity in the 
    outer jet region. The VLBI observations at 15\,GHz also revealed that
     knot R3 had its kinematic behavior similar to that of knot B4 observed 
     at 43\,GHz, while the kinematic behavior of  knot R1 and R2 was quite 
     different from that of knot R3;  
     (2) At both 43\,GHz and 15\,GHz a prominent arc-like structure was 
    detected, distributing over a very broad range of position angle from
    $\sim$--${120^{\circ}}$ to $\sim$--${30^{\circ}}$. A rapid position angle
     swing of $\sim$${70^{\circ}}$ between knot B4 and knot B6 in $\sim$1.3 years
     might be regarded as a clue for the existence of a double-jet structure
     in 3C454.3, indicating that knot B4 and knot B6 were ejected from their
     respective jet; (3) The recurrent trajectory patterns found in the
     knot-pairs B4/K09, B6/K10 and B2/K14 may be regarded as favorable 
     evidence to suggest some periodic behavior in knots' kinematics induced 
     by the jet-nozzle precession with a precession period
     of $\sim$10.5\,yr and the possible existence of a precessing common
     trajectory. Such kind of periodic recurrence of curved trajectory patterns 
     seems to be important signatures for recognizing nozzle precession and
     determining precession periods for blazars; (4) Knot B4 produced 
     two radio bursts at 43\,GHz: one 
     occurred near the core and the other in the outer jet region. Similarly,
     knot R3 also  produced two radio bursts at 15\,GHz, but at different core
     distances. Both the bursts can be interpreted in terms of Doppler-boosting
     effect. This similarity observed at diffrent frequencies might indicate
     the stability of the track patterns for knot B4 and knot R3. \\
       Our precessing nozzle scenario is based on the two assumptions:
    (1) Superluminal components are ejected from a jet-nozzle at 
    corresponding precession phases when the jet-nozzle precesses. Here for 
    3C454.3, the precession period is found to be 10.5 years; (2) Superluminal
    components move along their common precessing tracks in the
     innermost jet regions with a transition at certain core-distances in the
    outer-jet regions from precessing common tracks to their own individual
    trajectories.\\
    However, these assumptions may imply to introduce severe constraints on
    the precessing nozzle secnario when it is applied to investigate the
    VLBI-kinematics of superluminal components in blazar 3C454.3 (Qian et al.
    \cite{Qi21}).\footnote{Also in 3C345 (Qian \cite{Qi22a}, \cite{Qi22b}); 
    OJ287 (Qian \cite{Qi18b}) and 3C279 (Qian et al. \cite{Qi19}).} That is,
     these assumptions may be "unavoidable"  to require
     double precessing-nozzle structures existing in these blazars. Otherwise,
    if "single jet-nozzle" structures are assumed for these blazars,
    there would be no unified nozzle-precession to be explored for
    consistently delineating the kinematics of their superluminal components
    as a whole. One would have to deal with the kinematics of many superluminal
    components which are independent of each other. Thus it would be difficult
    to clarify their respective characteristics and connections to the central
    engine (the black hole/accretion disk system in its nucleus) for 
    the components as a whole. \\
      In contrast, for quasars (e.g., B1308+328, PG1302-102 and NRAO 150) the
    precessing jet-nozzle scenario with a single jet structure may be
     applicable to determine their precession periods (Qian et al.
    \cite{Qi18a}, \cite{Qi17}; Qian \cite{Qi16} and  Qian \cite{Qi23}).\\
    Therefore, at present, the assumption of "double jet-nozzle structure" for 
    blazar 3C454.3 (also for blazars  
    3C279, 3C345 and OJ287)  may be regarded as a "working hypothesis",
    although it has initiated some physically meaningful results and there 
    are some observational clues in favor of this suggestion.
     \footnote{MHD theories of relativistic jets also provide some arguments
     for the possible existence of
     double-jet structure in binary black hole systems (Artymovicz \& Lubow
     \cite{Ar96}, Artymovicz \cite{Ar98}, Shi et al. \cite{Sh12},
    Shi \& Krolik \cite{Sh15}).}\\ 
    In summary,  we have applied our precessing nozzle scenario to study the
    kinematics and flux evolution of superluminal components in a few QSOs
     and blazars, determining their precession periods and precessing common 
    trajectory patterns. For quasars B1308+326, PG1302-102 and NRAO150 only
    single jet structure has been assumed, while
    for blazars 3C279, 3C345, 3C454.3  and OJ287 double-jet structures are
    assumed. In all these cases the flux evolution of superluminal components
    can be interpreted in terms of their Doppler-boosting effect, although
    their intrinsic variations on shorter time-scales induced by the evolution
    of superluminal knots (superluminal plasmoids or relativistic shocks)
    need to be taken into account. The combined effects of Doppler-boosting and 
    intrinsic variation in the flux evolution of superluminal components 
    have also been found in blazar 3C345 (especially for its knots C9 and C10;
     Qian \cite{Qi22a}, \cite{Qi22b}).\\
    Generally, the assumption of precessing common trajectory may be applicable,
    but the model-derived patterns and their extensions
    are quite different for different superluminal knots 
    (Qian et al. \cite{Qi21} and references  therein). 
   Higher resolution VLBI observations are needed to show if this assumption
    is still valid within core separations $<$0.1\,mas.\\
    Theoretically, this assumption may be based on the theoretical works of
     relativistic magnetohydrodynamics for jet formation:  
     magnetic effects in acceleration/collimation zone of relativistic jets 
    are very strong, forming some very solid magnetic structures and 
    controlling the trajectory patterns of moving 
    superluminal knots ejected from the jet nozzle (e.g., Vlahakis 
    \& K\"onigl \cite{Vl04}, Blandford \& Payne \cite{Bl82},
     Blandford \& Znajek \cite{Bl77}, Camenzind \cite{Cam90}  and 
     references therein). Thus all the superluminal components in a blazar can
    move along the precessing common trajectory  if the jet nozzle is 
    precessing.\\
  \begin{acknowledgement} 
  We would like to thank J.W.~Qian for her help with the preparation
  of the Figures.  
  \end{acknowledgement}
   
  \newpage
   \begin{appendix}
   \section{Flux evolution of superluminal knot R3 observed at 15GHz}
    3C454.3 has a very complex structure at 15\,GHz as like at 43\,GHz. 
    The most distinctive and prominent feature found by the VLBI-observations
    at 15\,GHz during the period (1995 Jul.29--2010 Nov.20; Britzen et al.
    \cite{Br13}, Qian et al. \cite{Qi14}) was the evolving arc-like structure
    around the core (Figs.A.1 and A.2), which was constituted by the knots
    (R1, R2, R4 and R3) moving along different curved trajectories and 
   distributing in an area delimited by $r_n{\simeq}$[2.5, 3.5]\,mas and 
   PA${\simeq}$[--$40^{\circ}$, --$110^{\circ}$]. Such an arc-like structure
   observed at 15\,GHz is evry similar to that observed at 43\,GHz 
   (cf. Fig.1 in the text), but at different distances to the core.\\
    The kinematic behavior of its eight superluminal components (R1, R2, 
    R3, R4, A, B, C and D) has been analyzed (Qian et al. \cite{Qi14}). 
    It was found that their kinematic features (including the projected 
    trajectory and the core separation and coordinates versus time) could 
    be well model-fitted and interpreted in terms of our precessing jet-nozzle 
    scenario. Their apparent velocity/viewing angle and bulk-Lorentz/Doppler
    factor as function of time were derived.\\
    In addition, knot R3 was tracked by the 15\,GHz VLBI observations from
    core distance ${r_n}\sim$0.7\,mas to $\sim$3.7\,mas and a prominent 
    trajectory-curvature at core distances $\sim$2--3.84\,mas 
    (Figs.A.4. and A.5) was detected. Interestingly, a radio flare was 
    observed during $\sim$2005--2012 to be closely associated with this 
    curvature (Fig.A.7). This phenomenon could be related to the convergence 
    and re-collimation of the 'mini-jet (or beam)' associated 
     with the knot (R3), resulting in its re-acceleration and an 
    increase in its bulk Lorentz factor (cf. Figs. A.2--A.6).\\
     It should be noted that knot R3 produced two flares at 15\,GHz: one 
    in $\sim$1995 (only part of it was measured) and 
    the other during $\sim$2005--2012. So the flaring behavior  of knot R3 at
    15\,GHz is quite similar to the flaring behavior of knot B4 at 43\,GHz,
    which also produced two flares with the second one associated with the
    trajectory-curvature at $r_n{\simeq}$0.7--1.2\,mas (during 
    $\sim$1999.8--2001.4; cf. Figs.\,5--8 in the text).
    \footnote{Thus we would presume that knot R3 could be ascribed to the 
    jet-A in the double-jet scenario, as the knot B4 is.}\\
    \subsection{Knot R3: model-fitting of kinematics at 15\,GHz}
    The kinematics observed at 15\,GHz for knot R3 has already been well
     model-fitted in terms of our precessing jet-nozzle scenario in Qian et al.
    (\cite{Qi14}), where the precessing common trajectory pattern and its
    ejection epoch ($t_0$=1992.0 corresponding to a precession phase 
   $\phi$=5.75\,rad) were determined. We shall utilize these results and
   more carefully  perform the model-fits of its kinematic features 
    (trajectory $Z_n(X_n)$, core distance $r_n(t)$, coordinates $X_n(t)$ 
    and $Z_n(t)$) in order to correctly derive its bulk Lorentz factor 
    $\Gamma(t)$ and Doppler factor $\delta(t)$ for explaining its flux-density
    evolution.\\
     \begin{figure*}
     \centering
     \includegraphics[width=7cm,angle=0]{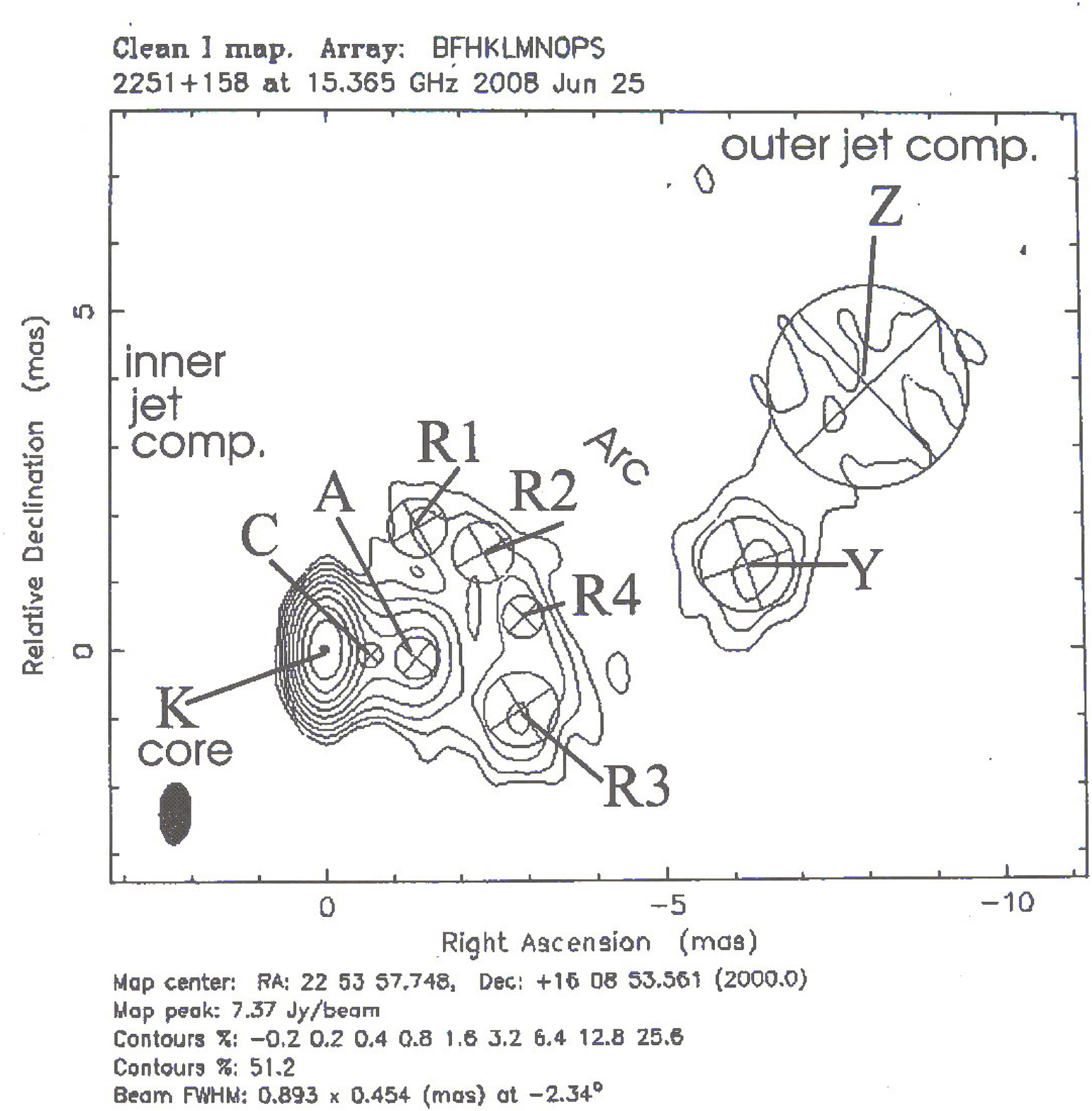}
     \caption{15GHz-image of 3C454.3 measured on 25 June 2008. Knots R1,R2,
      R3 and R4 distribute along an arc-like structure in a fan area delimited
      by position angle=[--$45^{\circ}$, --$103^{\circ}$] and core separation
     $r_n$=[2\,mas, 4\,mas]. In comparion with the 43\,GHz map (Fig.1 in the 
    text), knot R3 could be associated with the jet-A in the double-nozzle
     scenario proposed for interpreting the 43\,GHz kinematics. Thus the 
     second radio burst observed at 15\,GHz during 2005-2011 could originate
     from similar physical causes as the second burst observed at 43\,GHz
      during 1999.8--2001.2.}
     \end{figure*}
     \begin{figure*}
     \centering 
     \includegraphics[width=4.5cm,angle=-90]{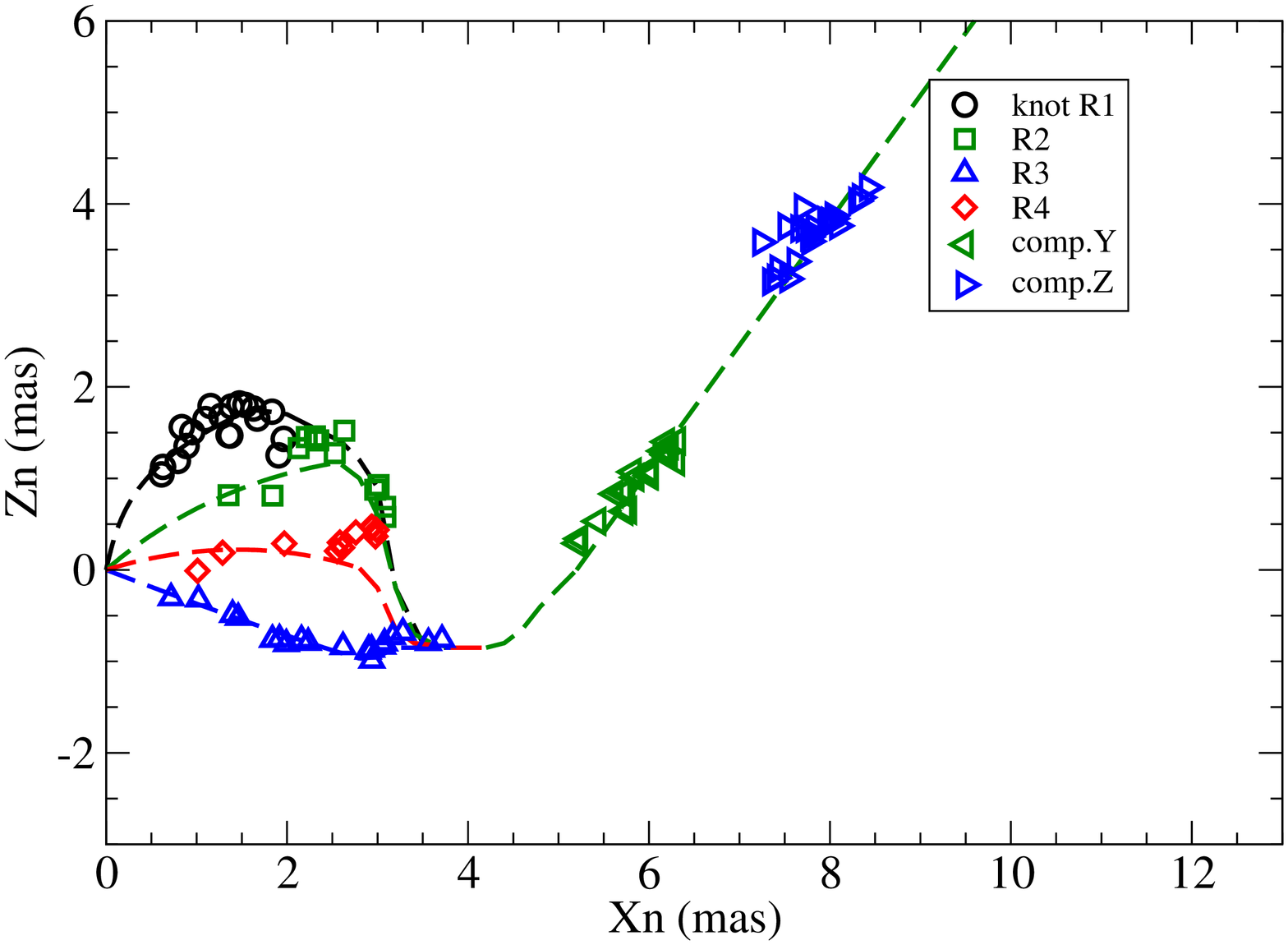}
     \includegraphics[width=4.5cm,angle=-90]{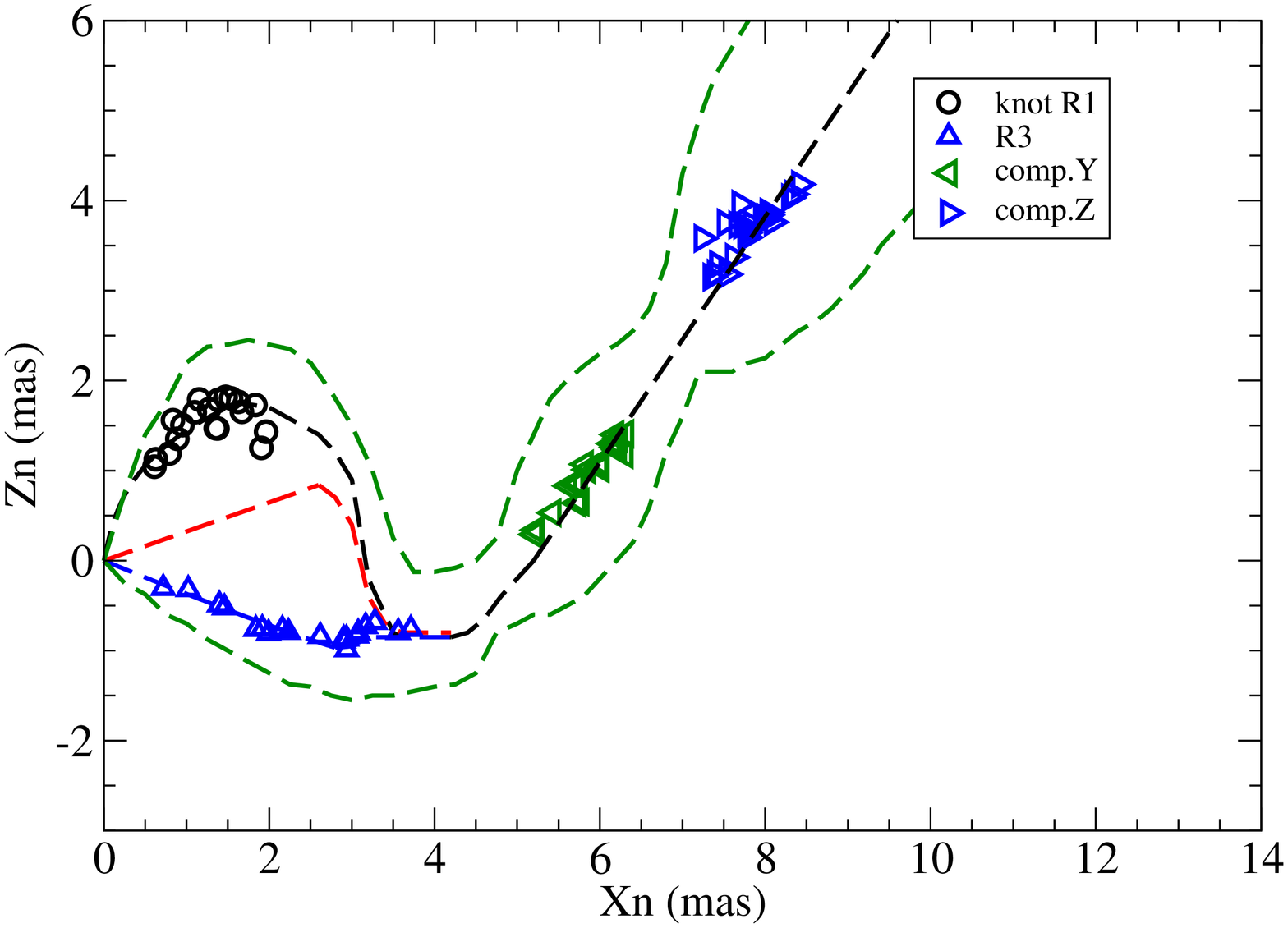}
     \caption{A conceptual sketch illustrating the apparent 
    convergence of the jet-trajectories and the formation of the arc-like 
    structure (Qian et al. \cite{Qi14}). Near the position 
    ($X_n$, $Z_n$)=(3.0\,mas, --1.0\,mas) knot R3 passed through a 
    convergence/recollimation region and then it moved toward the direction of 
    knot Y, producing a radio burst. In  contrast, knots R1 and R2 
    had no extensions along the position angles of 
    $\sim$--$40^{\circ}$ -- $\sim$--$50^{\circ}$. The green dashed lines 
    in the right panel indicate the jet boundaries
     defined by the sizes of the superluminal components.}
     \end{figure*}
     \begin{figure*}
     \centering
     \includegraphics[width=5.6cm,angle=-90]{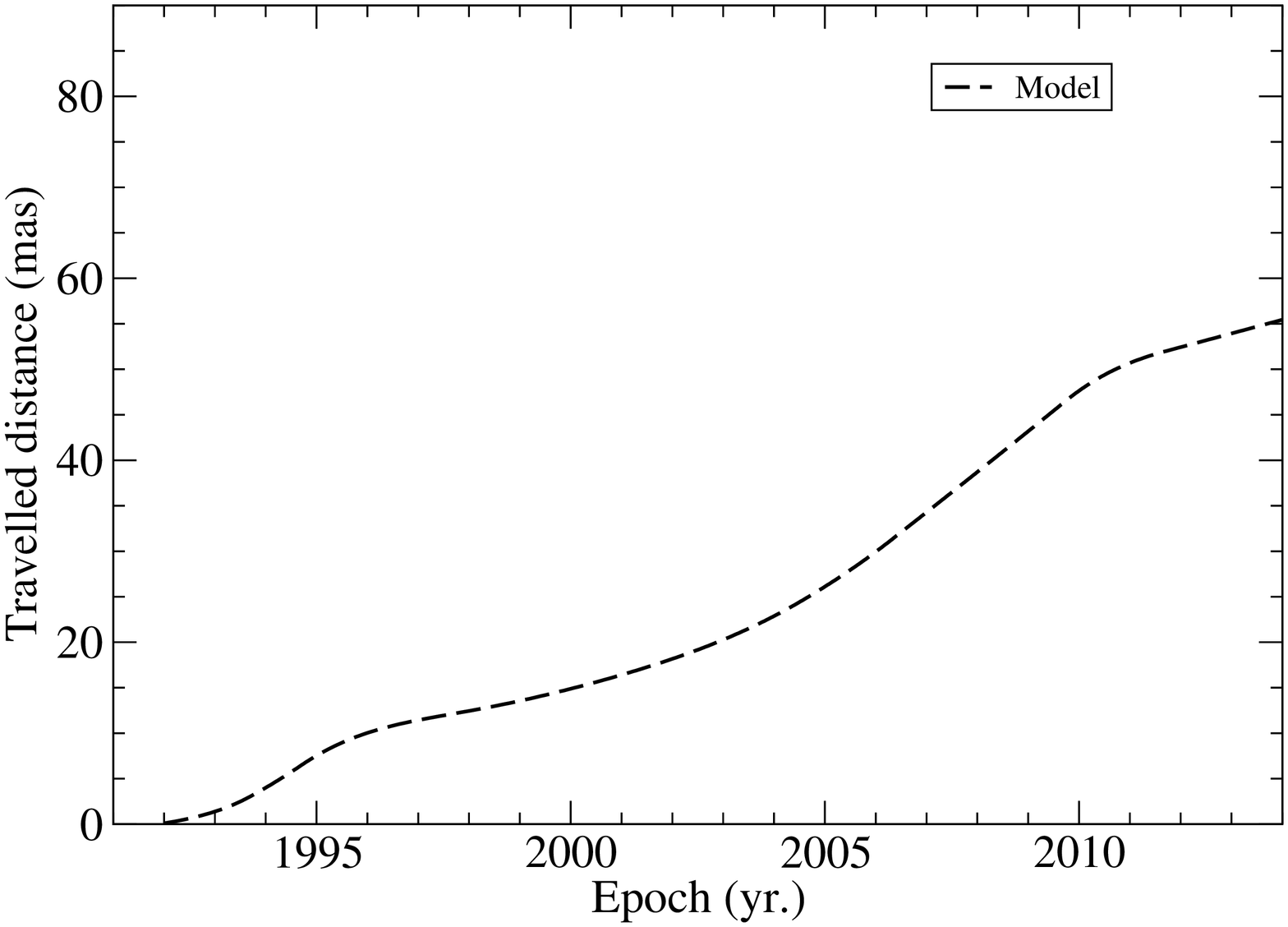}
     \includegraphics[width=5.6cm,angle=-90]{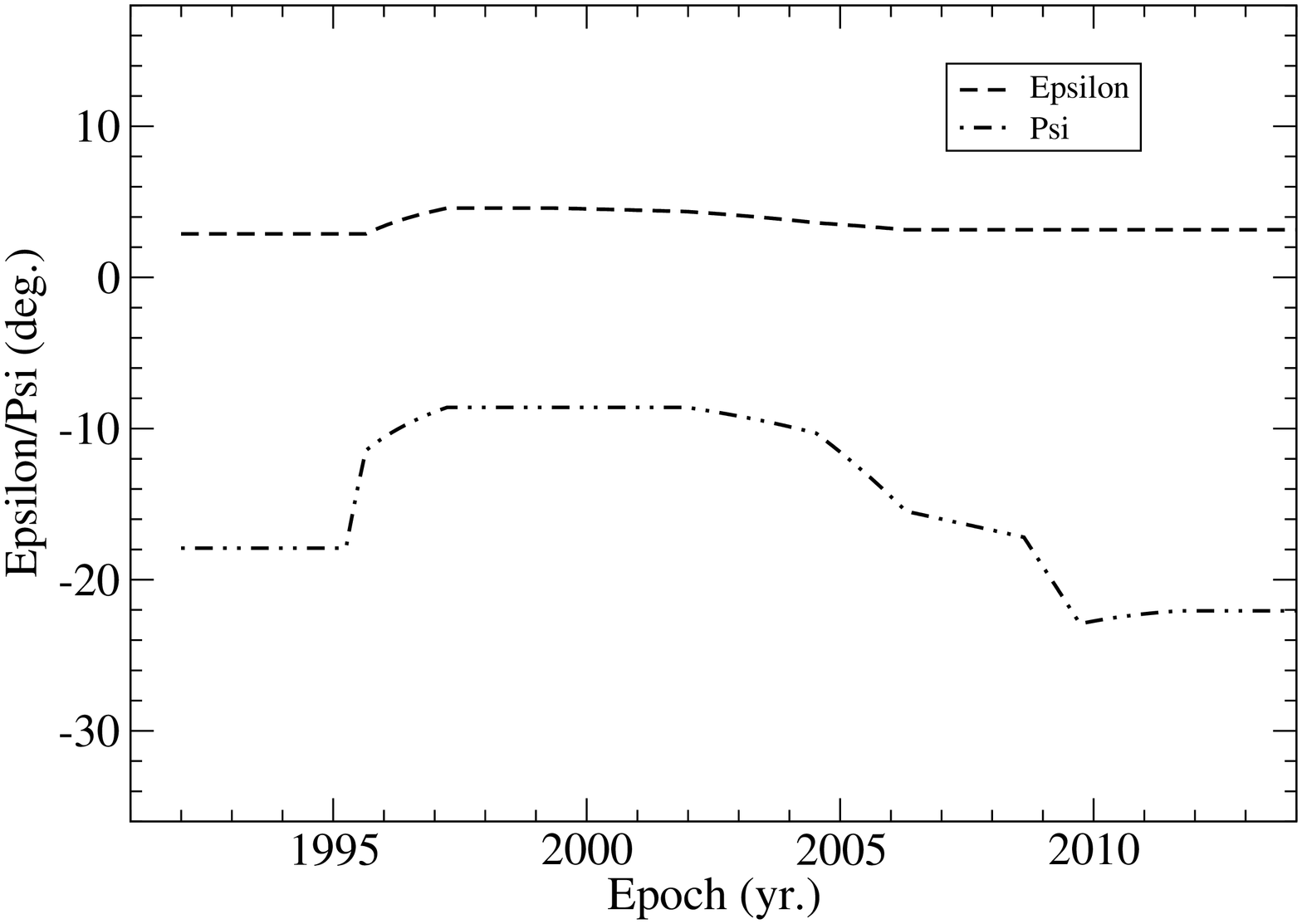}
     \caption{Knot R3: the model-simulated traveled distance $Z(t)$ along
     the Z-axis (left panel) and the modeled curves of parameters $\epsilon(t)$
     and $\psi(t)$, which define the plane where the jet-axis locates. In the
     innermost jet region (core separation 
    $r_n{\leq}$0.58\,mas (or before 1995.25)
    knot R3 moved along the precessing common trajectory with its precession
     phase $\phi$=5.75\,rad and corresponding ejection time $t_0$=1992.0.
    After 1995.25 $\psi$ changed largely and R3 started to move along its own
    individual track. During 2005--2011 parameter $\psi$ decreased quickly,
    inducing an oscillating curvature in its trajectory (Fig.A.4).}
     \end{figure*}
     \begin{figure*}
     \centering
     \includegraphics[width=5.6cm,angle=-90]{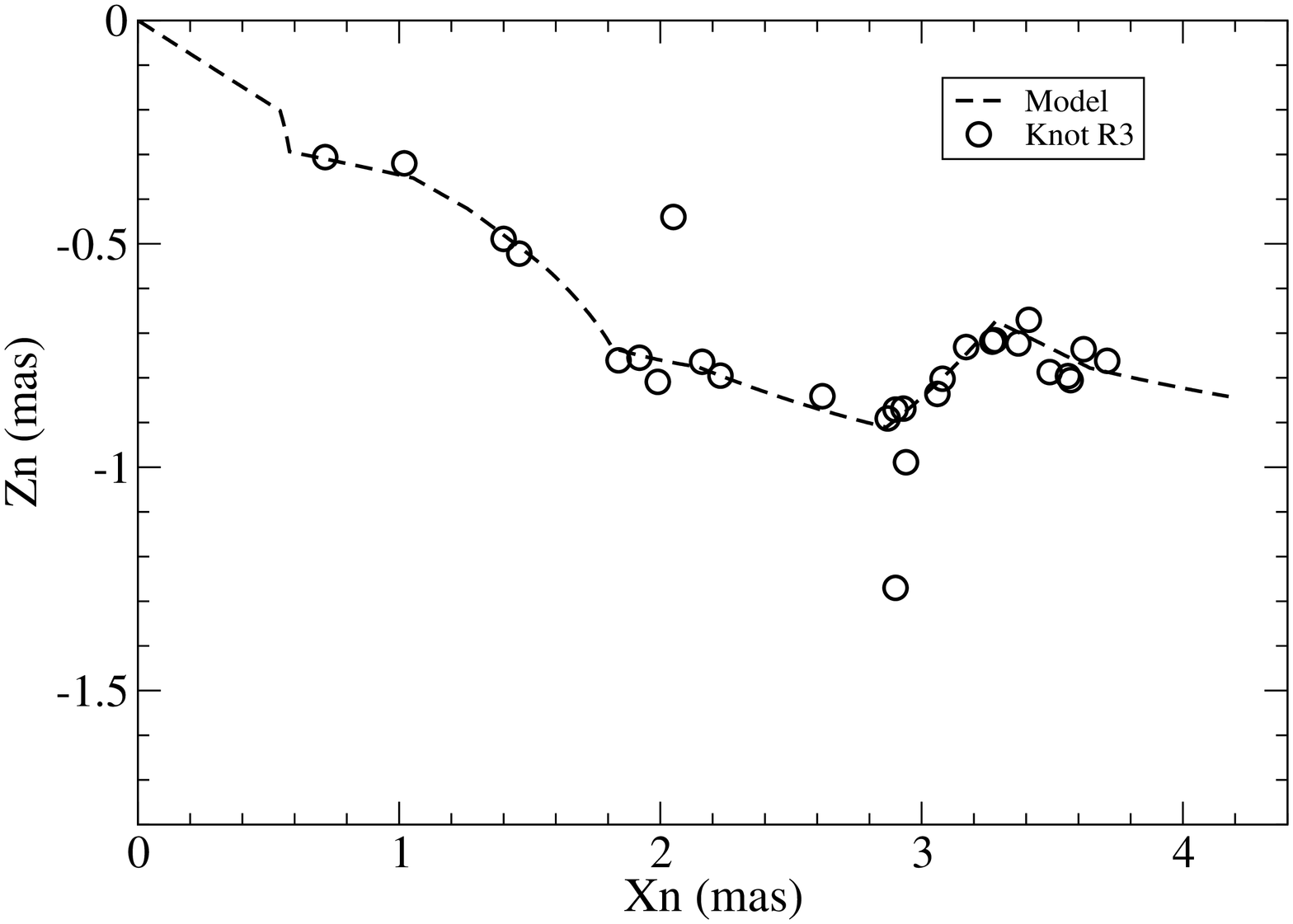}
     \includegraphics[width=5.6cm,angle=-90]{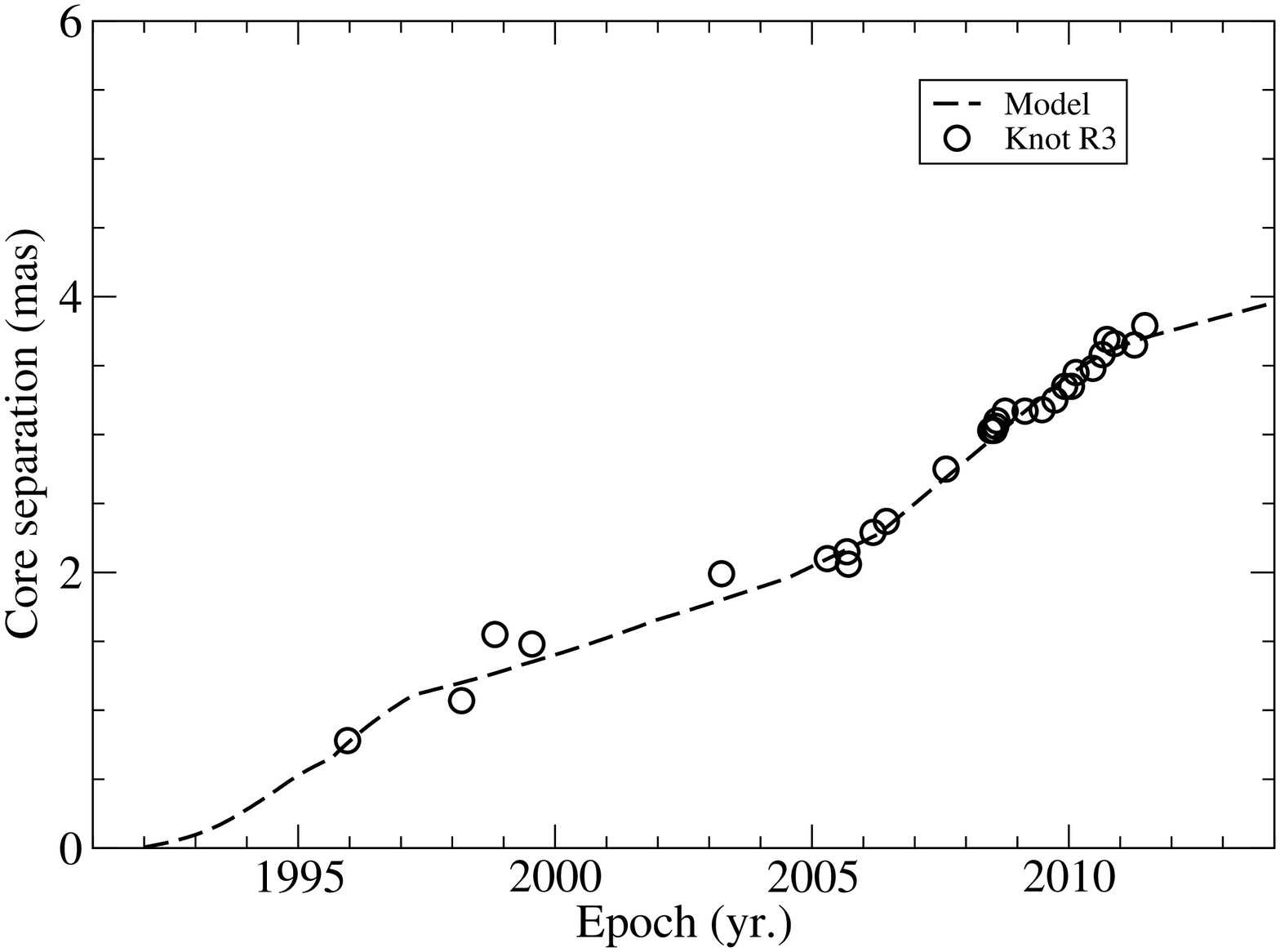}
     \caption{Knot R3: the model-fits of the projected trajectory $Z_n(X_n)$
     (left panel) and the core separation $r_n(t)$. The entire trajectory 
     extending to $r_n{\sim}$3.8\,mas (Z=408.3\,pc) was well fitted, 
     especially the oscillating curvature in the range of 
     $X_n{\sim}$[2--3.8]\,mas. Within 
     $X_n{\sim}$0.54\,mas (Z$\leq$8.3\,mas=63.8\,pc; or before
      $\sim$1995.25) knot R3 moved along the precessing common trajectory.
      Unfortunately no VLBI-observations were performed then. After 1995.25
     knot R3 started to move along its own individual track, deviating from
     the precessing common trajectory.}
     \end{figure*}
     \begin{figure*}
     \centering
     \includegraphics[width=5.6cm,angle=-90]{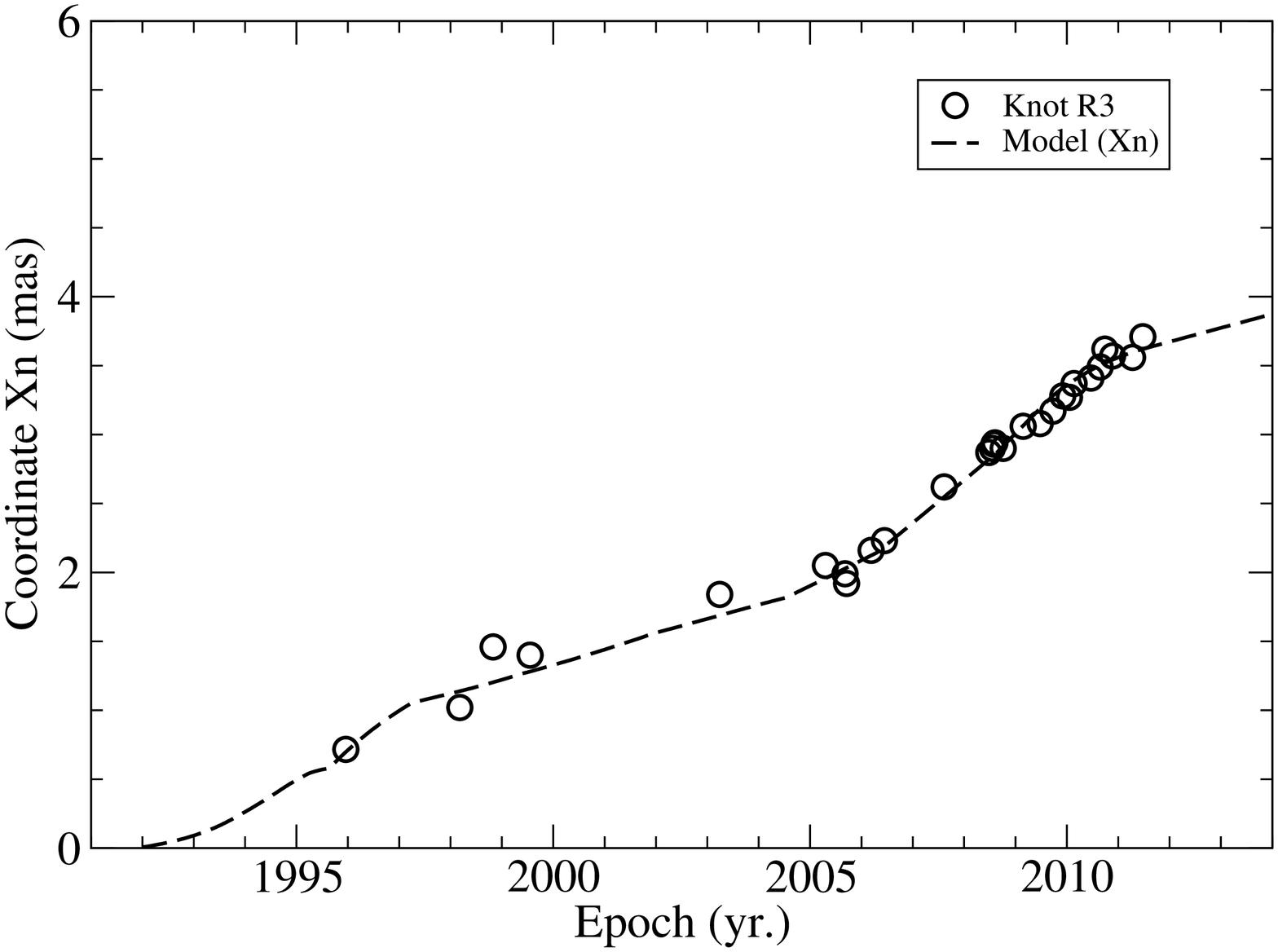}
     \includegraphics[width=5.6cm,angle=-90]{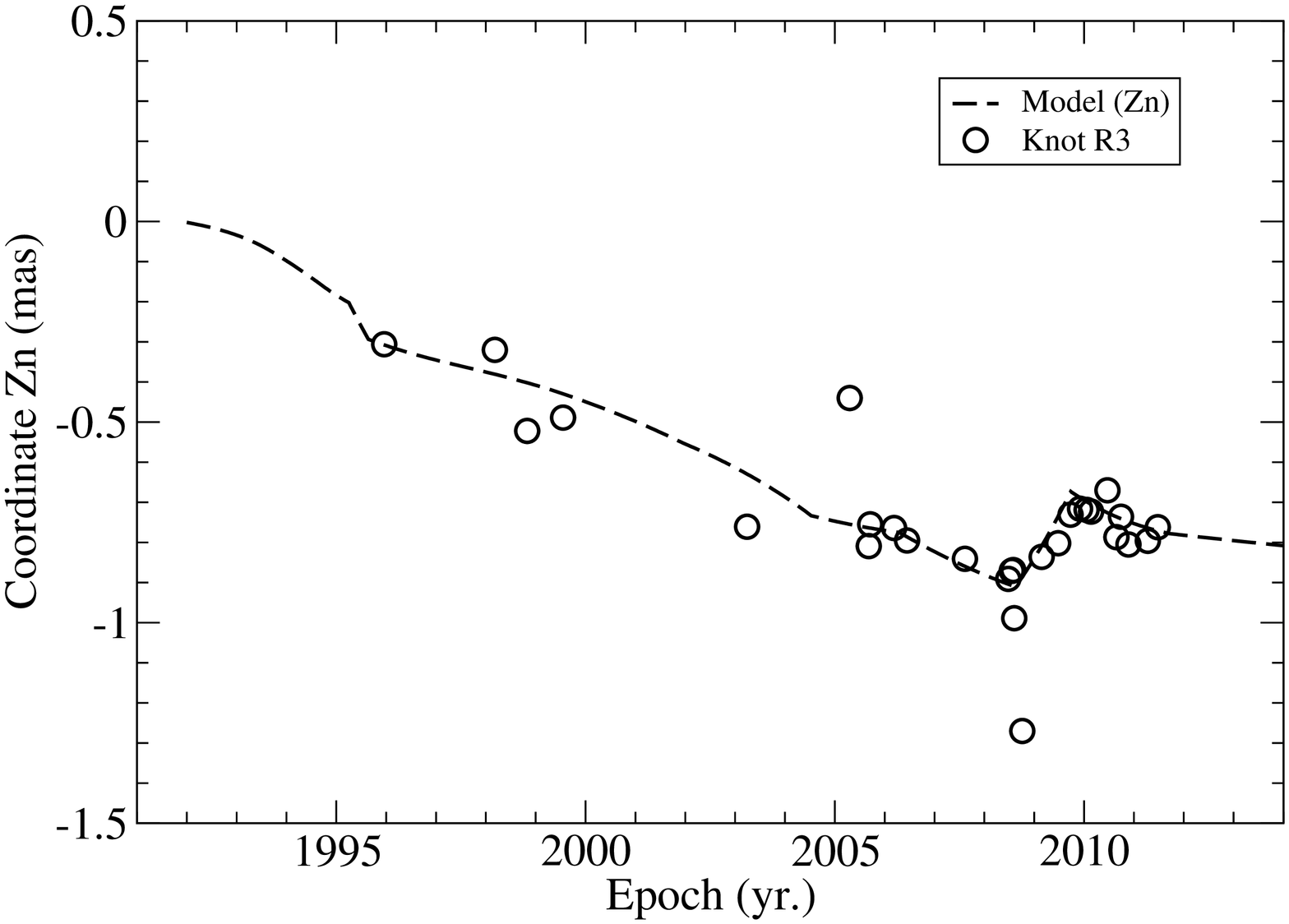}
     \caption{Knot R3: the model-fits of the coordinates $X_n(t)$ and $Z_n(t)$
      which were well fitted entirely, especially the oscillating curvature
     during 2005--2011 (right panel).}
     \end{figure*}
     \begin{figure*}
     \centering
     \includegraphics[width=5.6cm,angle=-90]{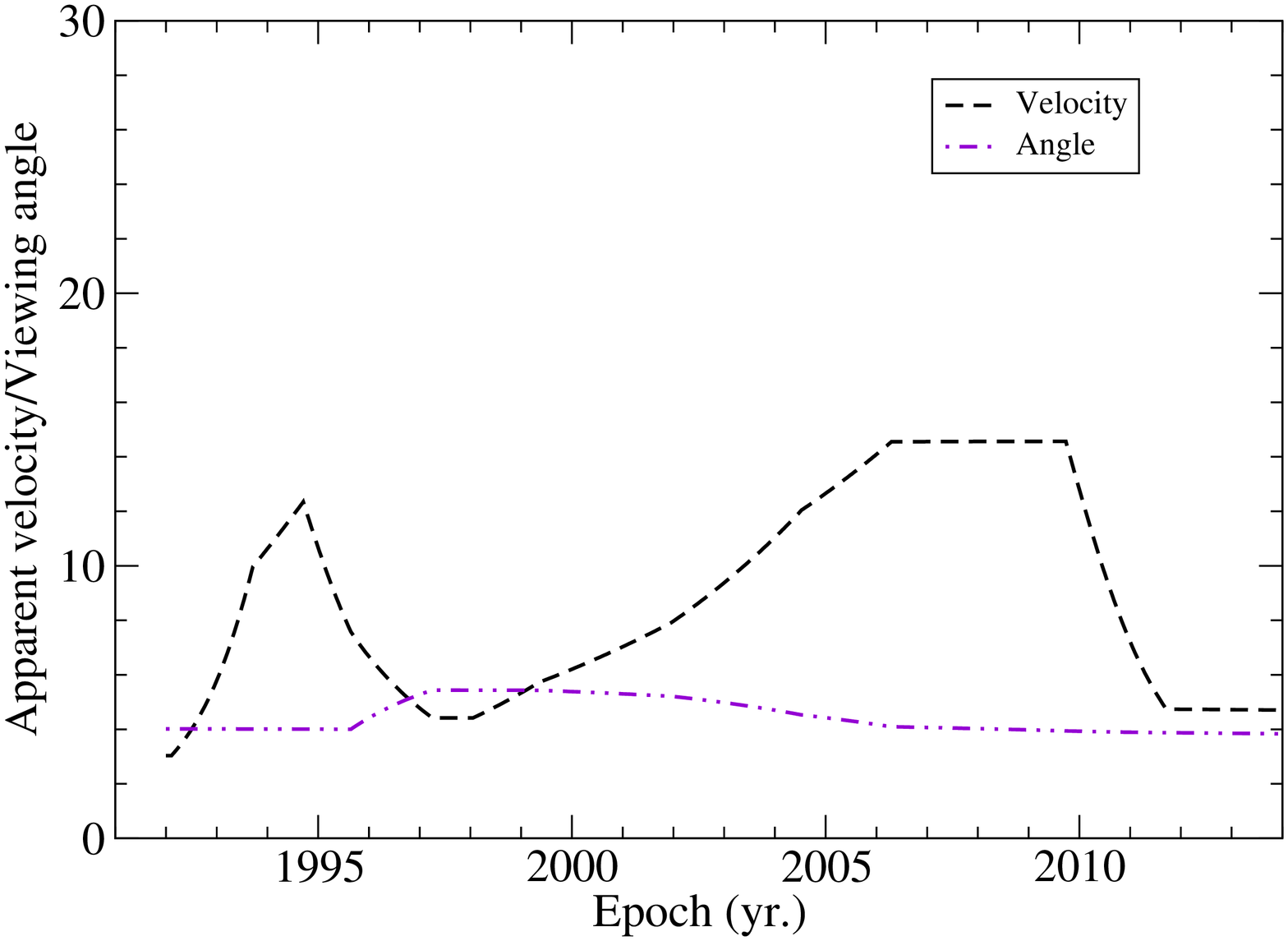}
     \includegraphics[width=5.6cm,angle=-90]{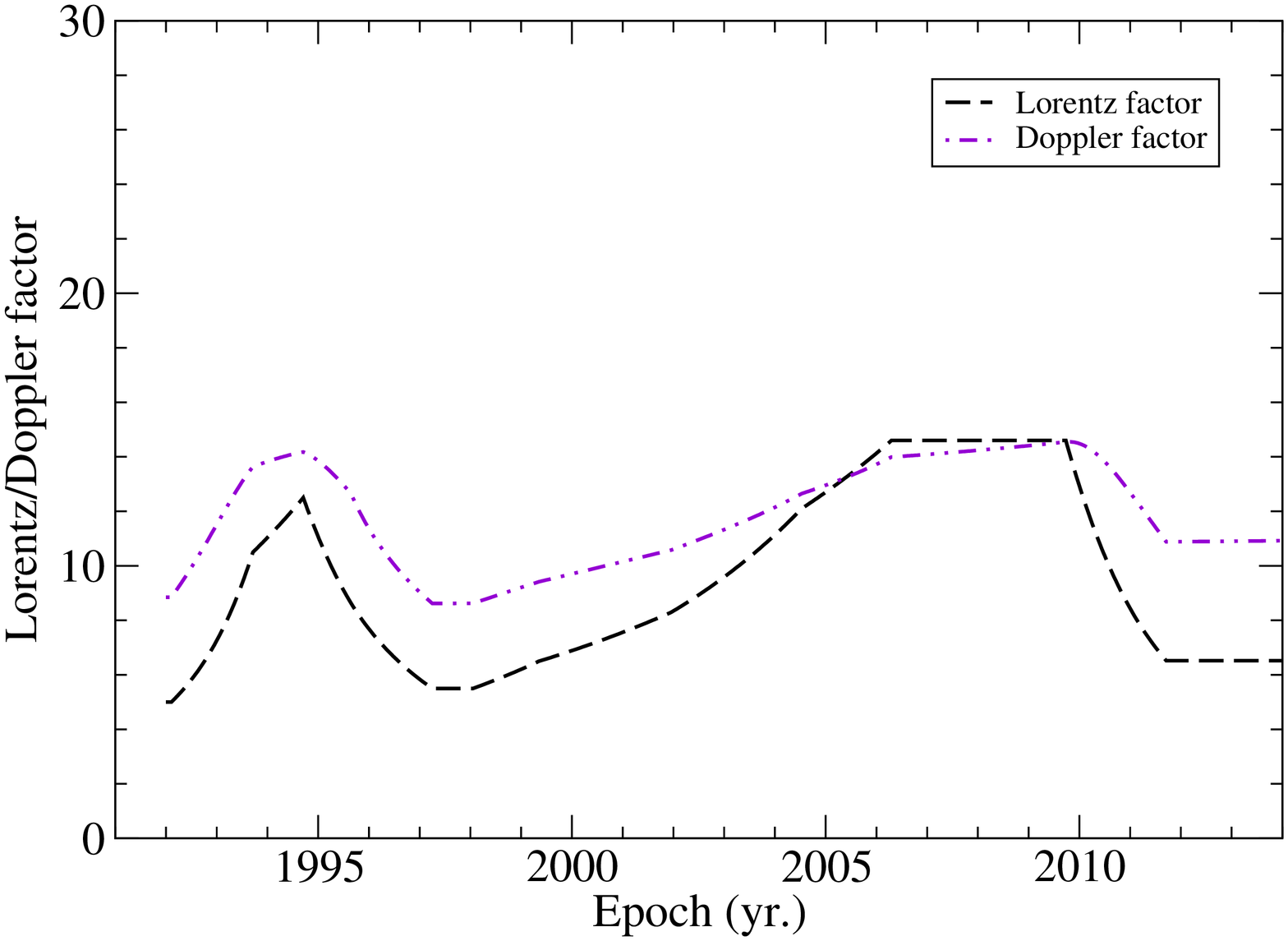}
     \caption{Knot R3: the model-derived apparent velocity $\beta_{app}(t)$
     and viewing angle  $\theta(t)$ (left panel) and the model-derived 
     bulk Lorentz factor $\Gamma(t)$ and Doppler factor $\delta(t)$. 
     $\beta_{app}(t)$, $\Gamma(t)$ and $\delta(t)$ all have a double-bump
    structure: at the first peak (epoch 1994.70): $\delta_{max}$=14.2,
     $\Gamma_{max}$=12.5 and $\beta_{app,max}$=12.3. At the second peak 
     (epoch 2009.80), $\delta_{max}$=14.6, $\beta_{app}$=14.1 and
   $\Gamma$=14.2, while at 2009.73 $\Gamma_{max}$=14.6, $\beta_{app,max}$=14.6 
   and $\delta$=14.5. The viewing angle $\theta(t)$ varied in the range
      [$3.9^{\circ}$(1992.0)--$5.4^{\circ}$(1997.2)--$3.9^{\circ}$(2011.0)].} 
     \end{figure*}
     The model-fitting results are presented in Figs.\,A.3--A.7.
     In Fig.A.3 are presented the traveled distance Z(t) along the jet-axis 
     (left panel) and the model-derived curves for parameters $\epsilon(t)$ 
     and $\psi(t)$ (right panel). Within core 
    separation $r_n$=0.58\,mas (corresponding to Z$\leq$8.3\,mas=63.8\,pc,
    or before 1995.25), [$\epsilon$, $\psi$]=[$2.88^{\circ}$, --$17.9^{\circ}$],
      knot R3 moved along the precessing common trajectory with its precession
      phase $\phi$=5.75\,rad and ejection time $t_0$=1992.0. After 1995.25
     parameter $\psi$ changed largely and R3 started to move along its own 
     individual track, deviating from the precessing common trajectory. During 
     the time-interval $\sim$2005--2011 ($X_n{\simeq}$2--3.8\,mas) changes in
      parameter $\psi$ resulted in an oscillating curvature in its trajectory
     (Fig.\,A.4, left panel). \\
    \begin{figure*}
     \centering
     \includegraphics[width=5.6cm,angle=-90]{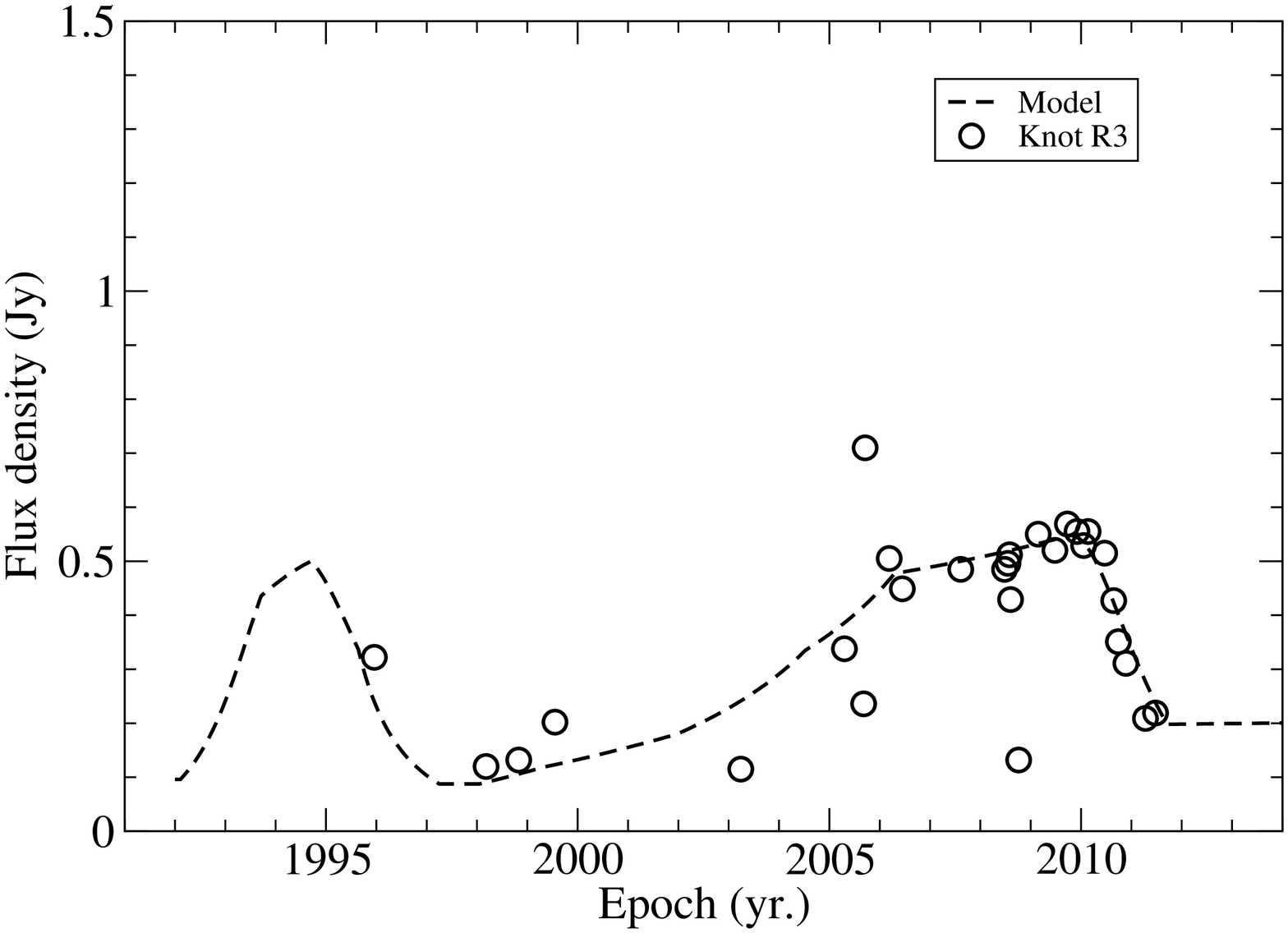}
     \includegraphics[width=5.6cm,angle=-90]{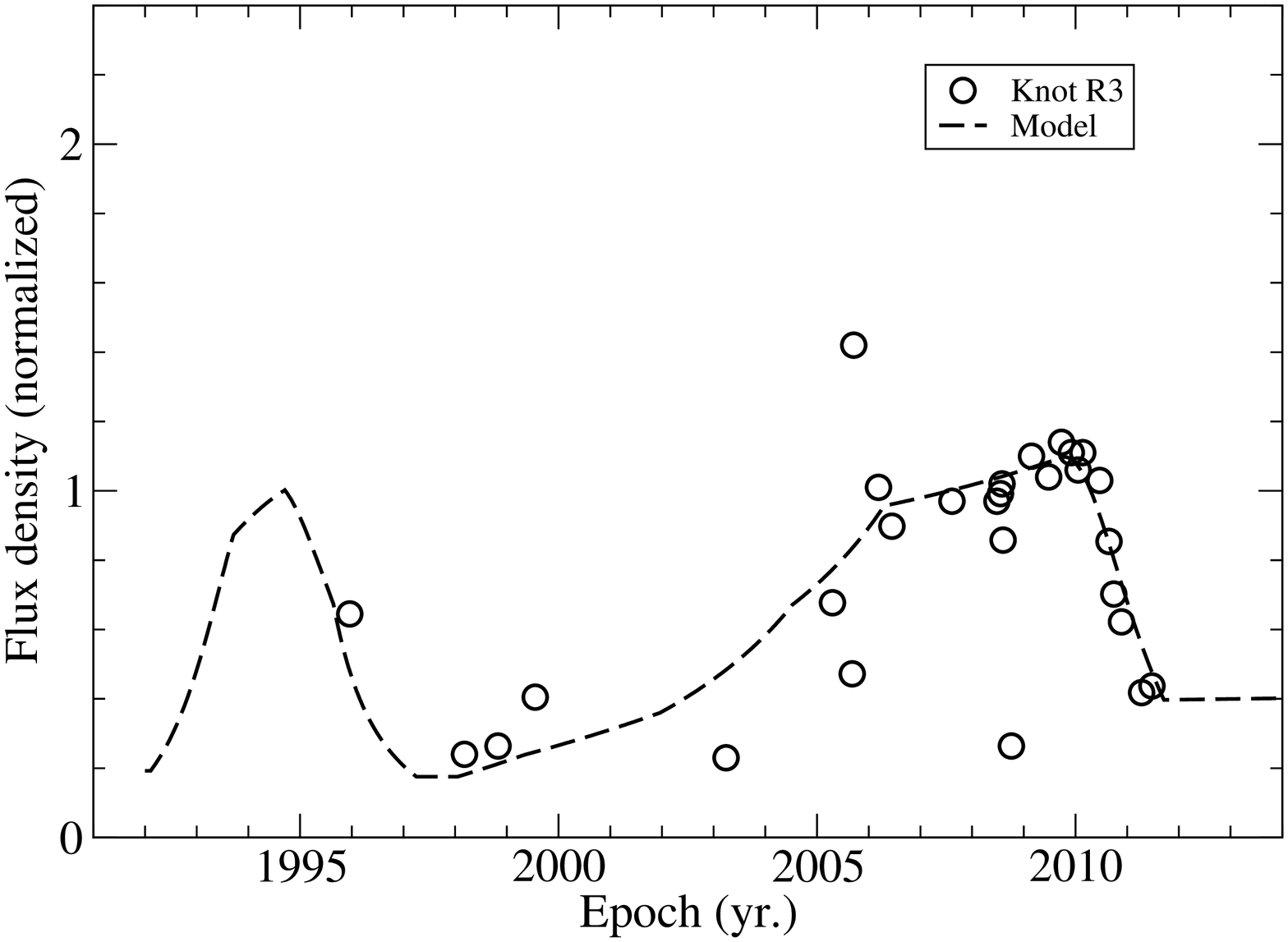}
     \caption{Knot R3. Left panel: the model-fit of the
     15\,GHz light curve in terms of the Doppler-boosting effect with
     the peak flux density being 0.50\,Jy and 0.55\,Jy for the two radio bursts,
      respectively. The intrinsic flux density $S_{int}$=46.6$\mu$Jy.
     Right panel: the model-fit of the light curve normalized by the peak flux
   density of the first burst  was well fitted in terms of the Doppler-boosting
      profile $[\delta(t)/\delta_{max}]^{3+\alpha}$ with an assumed value for
      $\alpha$=0.5.}
     \end{figure*}
        The model-fits of the trajectory $Z_n(X_n)$, core separarion 
     $r_n(t)$, coordinates $X_n(t)$ and $Z_n(t)$ are shown in Figs.\,A.4 and 
     A.5, respectively. It can be seen that all the kinematic features were 
     precisely fitted with the entire trajectory extending to core 
     separation of $\sim$3.8\,mas and the oscillating curvature in the 
     trajectory at $X_n{\simeq}$2--3.8\,mas.\\
     The  model-derived apparent velocity $\beta_{app}(t)$ and viewing angle
     $\theta(t)$ are presented in Fig.A.6 (left panel). The model-derived
      bulk Lorentz factor $\Gamma(t)$ and Doppler factor $\delta(t)$
     are shown in Fig.A.6 (right panel). The model-derived curves for 
     $\beta_{app}(t)$, $\Gamma(t)$ and $\delta(t)$ all 
     have two-bump structures corresponding to the two radio bursts. 
     At the first bump (epoch 1994.7): $\delta_{max}$=14.2, 
     $\Gamma$=$\Gamma_{max}$=12.5 and $\beta_{app}$=$\beta_{app,max}$=12.3.
     For the second broad bump, at epoch 2009.80: $\delta_{max}$=14.6, 
     $\Gamma$=14.2 and $\beta_{app}$=12.1, while at epoch 2009.73 
     $\Gamma$=$\Gamma_{max}$=14.6, $\beta_{app}$=$\beta_{app,max}$=14.6 and
     $\delta$=14.54. The viewing angle $\theta(t)$ varied in the range of
     [$3.9^{\circ}$(1992.0)--$5.4^{\circ}$(1997.2)--$3.9^{\circ}$(2011.0)].\\
    \subsection{Knot R3: model-fit of 15\,GHz light curve}
     Since  the kinematics of knot R3 was successfully  model-fitted and its
     bulk Lorentz factor and Doppler factor versus time were precisely derived,
     the 15\,GHz light curve of knot R3 was well fitted in terms of its
     Doppler boosting effect, as shown in Fig.A.7. The model-fit of the 
     observed light curve is shown in the left panel with its peak flux density
     0.50\,Jy and  intrinsic flux density of 46.6$\mu$Jy. The light curve 
     normalized by the peak flux density of the first burst was well fitted by 
     the Doppler-boosting profile $[\delta(t)/\delta_{max}]^{3+\alpha}$ with a
     presumed value $\alpha$=0.5 as shown in the right panel. The peak flux 
    density of the second burst is modeled as 0.55\,Jy.\\
      It worths noting that the second burst (during $\sim$2005--2010) of knot
       R3 occurred in the outer jet-regions at core distances 
     $r_n{\simeq}$[2.0--3.8]\,mas, where its trajectory underwent an 
     oscillating curvature (Fig.A.4). This curvature was induced by the change
      in the parameter $\psi$ (Fig.A.3), which did not affect the variations 
      in the flux-density of knot R3. The second radio burst is mainly 
     induced by the Doppler boosting effect due to the increase/decrease in its
     bulk Lorentz factor and Doppler factor (Fig.A.6, right panel).\\
      The similarity of the morphological structures observed at 15\,GHz and
      43\,GHz for 3C454.3, and the occurrence of the second radio burst 
      observed in knot R3 (at 15GHz) and B4 (at 43GHz) clearly demonstrate that
      some similar physical environments, kinematic structures and conditions
      may exist on different scales in 3C454.3, resulting in similar kinematic 
      and emission properties for the superluminal components observed at 
      different frequencies, although they moved along different tracks with
     different speeds.

   \end{appendix}
 \end{document}